# Ionic conductance oscillations in sub-nanometer pores probed by optoelectronic control

Fanfan Chen[1#], Zonglin Gu[2#], Chunxiao Zhao[1], Yuang Chen[1], Xiaowei Jiang[1], Zhi He[2], Yuxian Lu[1], Ruhong Zhou[2,3] and Jiandong Feng[1]*

[1]*Laboratory of Experimental Physical Biology, Department of Chemistry, Zhejiang University, 310027 Hangzhou, China*

[2]*Institute of Quantitative Biology, College of Life Sciences and Department of Physics, Zhejiang University, 310027 Hangzhou, China*

[3]*Department of Chemistry, Colombia University, New York, NY10027, USA*

[#]*These authors contribute equally to the work*
*\*Correspondence should be addressed to jiandong.feng@zju.edu.cn*

**Abstract**

Ionic Coulomb blockade is one of the mesoscopic effects in ion transport revealing the quantized nature of ionic charges, which is of crucial importance to our understanding of the sub-continuum transport in nanofluidics and the mechanism of biological ion channels. Herein, we report an experimental observation and plausible theoretical reasoning of ionic conduction oscillations. Our experiment was performed under strong confinement in single sub-nanometer $MoS_2$ pores with optoelectronic control enabled for active tuning of pore surface charges. Under this charge control, we measured the ionic current at fixed voltages and observed multiple current peaks. Our analytical discussions and molecular dynamics simulations further reveal that the conductance oscillations in atomically thin nanopores may originate from the multi-ion interaction at the pore entry, particularly the electrostatic repulsion of ions external to the pore by ions bound inside the pore. Our work adds a further understanding of ionic Coulomb blockade effect under extreme confinement in atomically thin nanopores and paves the way for developing advanced ionic machineries.

**Introduction**

Ion transport plays a central role in the electrical signal transmission in living systems as well as in the operation of artificial nanofluidic devices. Nature has optimized highly selective and efficient transport systems using extreme confinement at the single ion level while artificial devices have just recently managed to approach such scales with precision nanofabrication capabilities and novel low-dimensional material platforms[1-5]. A number of novel physical effects are emerging in the experiments under strong confinement, evidencing that ion transport in the sub-nanometer (sub-nm) scale requires more universal descriptions, which might be out of the reach by classical continuum fluidics[6-8].

Many nanofluidic phenomena have successfully been explained in the continuum-based Poisson-Nernst-Planck (PNP) model, which treats Coulomb interactions between the ions at a mean field level. Recently, theoretical works have been proposed for capturing single ion transport in the framework of ionic Coulomb blockade[9-11] – an analogy to electronic Coulomb blockade[12]. Designing a sub-nm $MoS_2$ pore, in the absence of an external gate we observed signatures of ionic Coulomb blockade, with the features of non-Ohmic current voltage characteristics (*I-V*) and nonmonotonic conduction to pH control[13]. The origin of such effects in atomically thin nanopores remains mysterious and interesting because it is unclear how the concomitant electrostatic and dehydration effects are incorporated together in this confined system, owing to the lack of precise external gate control of the surface charge on the pore. In this work, we report the optoelectronic control of ion conduction in individual sub-nm pores which leads to a novel nonmonotonic phenomenology displaying ionic conductance oscillations. We further discuss a possible underlying transport mechanism in atomically thin nanopores.

**Optoelectronic control of ion transport in sub-nm pores**

To provide a first illustration of this process, we schematized the single ion transport process through a water-filling sub-nm $MoS_2$ pore and the experimental setup in **Figure 1a-c**. Since the $MoS_2$ pore is negatively charged and highly cation selective[5], we only show the conduction of potassium ions. As the $MoS_2$ pore is atomically thin, only about 0.65 nm, we could assume few potassium ions in the pore at each moment (**Figure 1a-b**). **Supplementary Figure 1** and **Supplementary Figure 2** show the detailed fabrication and characterization of sub-nm single layer $MoS_2$ pores. Both $MoS_2$ pore and graphene pore are considered in this work to show the universal appearance of this confined phenomenon in tiny and thin pores while the $MoS_2$ pore is used in the experiments due to the straightforward implementation of surface charge control. As shown in **Figure 1c**, active control over the surface charge of $MoS_2$ pores allows for the external gate modulation, acting as an additional dimension for controlling ion transport. When an electric field is applied across the membrane, ion transport through the pore will be driven and produces measurable current.

It has been recently demonstrated that the surface charge of nanopores can be precisely controlled by using a light based optoelectronic method[14,15]. The presence of a ~1.8 eV band gap for monolayer $MoS_2$[16] allows for the modulation of its charge carrier density and for tuning of the surface charge of MoS$_2$ nanopores using photoelectric effects[15]. We employ this approach for actively modulating the surface charge of MoS$_2$ pores using light as the external gate for this system (**Figure 1c**). Compared to the alternative electronic gate, optoelectronic control of surface charge by the integration of light illumination is straightforward to implement and does not directly introduce crosstalk with the ionic conduction channel. The optical beam profile and power density characterization of used laser sources (488 nm, 532 nm and 785 nm) are shown in **Supplementary Figure 3.** As presented in **Supplementary Figure 4**, an increase of ionic current is found upon application of light illumination. Ionic current noise spectral analysis in **Supplementary Figure 5-7** identifies the effective optical modulation, as evidenced by the increase of noise signals in the low frequency band[17]. In order to characterize how the light affects the surface charge of MoS$_2$ pores, we performed conductance measurements (**Supplementary Figure 8**) and ion selectivity measurements (**Supplementary Figure 9**) on MoS$_2$ pores. In the continuum range (pore size > 2 nm), the charge values could be extracted from experimental data using conductance-molarity relation at the surface regime[5,15]. Large pore data shown in **Supplementary Figure 8** results in an estimated charge density of 37.4 mC/m$^2$ that can be assigned to the corresponding light power density. For a MoS$_2$ pore, since the terminated pore edge atoms are responsible for preferentially carrying the surface charges, the sub-nm pores could contain higher charge density than those of larger pores. In addition, **Supplementary Figure 9** presents the concentration gradient measurements for a 2 nm MoS$_2$ pore where the reversal potential is increased from - 59.1 mV to -92.2 mV by the application of light (19.7 W/cm$^2$), meaning a significant increase of ion selectivity which we attribute to the effective optoelectronic modulation of pore charges on MoS$_2$ pores[15].

**Ionic conductance oscillations in MoS$_2$ pores**

Ion conduction modulation experiment is performed on sub-nm MoS$_2$ pores displaying non-linear *I-V*s (**Figure 1d**). By recording the ionic current at fixed transmembrane biases as a function of illuminating light power, current peaks occur at certain power densities (**Figure 1e**). Each optical modulation period is performed for at least two cycles to know whether these current peaks are actively controlled by the light illumination. Repeated recordings in **Figure 1f** indicate the oscillatory patterns are further recoverable.

Notably, although the shape of the oscillatory patterns does not allow for direct comparisons across different devices, it is recoverable in same devices in repeated modulation cycles. We have successfully performed full optical modulation experiments (excluding incomplete modulation, damaged devices during the operation, enlarged pore during experiments) on 33 sub-nm MoS$_2$ pore devices and observed distinguishable oscillations in 26 devices. There are also 6 devices displaying linear conductance increase under optical stimulation (**Supplementary Figure 10a**) and one device showing no response to light (**Supplementary Figure 10b**). The linear response may come from pore devices with poor photo-activity or weak surface charges. **Supplementary Figure 11** displays the reproduced oscillations from ten different

MoS$_2$ pore devices with qualitatively different response. This device to device variation (**Supplementary Figure 10** and **Supplementary Figure 11**) may originate from the rich possibilities allowing for heterogeneities of atomic arrangements, pore shape and charge of individual MoS$_2$ pores[18], which again points to the grand challenge of precise fabrication of artificial nanopores with well-defined atomic-scale configurations.

We further probed the current baseline fluctuations in response to the constant light power illumination. Results in **Supplementary Figure 12** and **13** present the long-term stability of the light modulated measurements which further suggest that these peaks are not from random baseline fluctuations or stochastic absorptions, but due to applied light control (More details in **Supplementary Section iv**). The stability and reversibility (**Figure 1f**, **Supplementary Figure 11** and **Supplementary Figure 12-15**) measurements show the initial pore conductance is almost recovered after the light modulation, indicating the nanopores are not enlarged considerably by the applied light owing to the arrived low power density, as estimated in **Supplementary Figure 3**. When the light power is out of the range that we currently report, it is possible that the laser could induce chemical modifications of the pore[19-21]. However, the activation of a photoinduced chemical reaction should be monotonic with the laser density which does not capture the oscillatory results. In repeated measurements (**Supplementary Figure 11**), the peak position does not always recover to the previous position even though the global features are matched. Such differences may originate from statistical photocharging process in sub-nm pores or additional laser induced thermal drift in the system.

To estimate the contribution from laser heating, we established a detailed heat transfer FEM model in solid and liquid (**Supplementary Section ii**) describing the experimental setup that we use. **Supplementary Figure 16** shows that the temperature increase due to laser heating is smaller than 4.5 K, far too small to induce the current changes given in the data. Finally, since the effect of temperature increase on the current is monotonic, increase in temperature cannot cause the phenomenon of current oscillations. Therefore, in agreement with other reports[14,15], we believe the laser induced heating is a minor effect in our system. Moreover, as the $MoS_2$ light interaction is highly specific to light wavelength[16], we compared the modulated experiments using different laser sources. In strong contrast to conductance oscillations observed for the 488 nm/532 nm laser modulation, the below-bandgap illumination at 785 nm light leads to a linear increase of ion current (**Supplementary Figure 17**) which we attribute to the laser heating of solution. This result further suggests that the current observations are of the optoelectronic origin.

The oscillatory current in sub-nm pore is in stark contrast to the linearized current increase in nanometer pores in response to optoelectronic modulation. When the pore size is increased, the light modulation leads to a linear increase of the current. The oscillatory to linear transition in a 1.4 nm pore is presented in **Supplementary Figure 14** and a linear modulation in a 15 nm pore is given in **Supplementary Figure 15,** showing a clear dependence of this observation on the pore size confinement. The linearized ion current vs light power density relation has also been reported in similar systems for pore sizes typically larger than 2 nm[14,15].

**Figure 2** displays the oscillatory ionic conduction in sub-nm pores under different ionic environment controls. Although the ionic current noise in the blockade regime gets slightly increased by the light illumination, the oscillatory results presented in **Figure 2** and **Supplementary Figure 18-19** from different devices are distinguishable from the baseline noise. Ionic concentration performs an important role, affecting surface charge, Debye screening and charge carrier density[9,11]. As shown in **Figure 2a-f**, we found the current oscillating ratio ($I_{min}/I_{max}$) decreased by about 10% when the concentration is changed from 0.1 M to 0.01 M owing to the surface charge determined transport in this system. The oscillatory strength is then compared for different ion valence solutions and the results presented in **Figure 2 g-l** appear to exhibit stronger oscillatory effects for ions with higher valence ($Al^{3+}$). However, the number of peaks found in **Figure 2 f-l** displays no clear difference for ions with varying valence.

To identify this valence effect, we have revisited the strong nonlinear $I$-$V$[13,22,23] in sub-nm pores. When the valence is increased from $K^+$ to $Al^{3+}$, **Figure 3a** reveals that the overall current suppresses accordingly while still maintaining strong nonlinear $I$-$V$. Differential conductance $dI/dV$ is plotted in **Figure 3b** for showing this non-*Ohmic* response. Strikingly, in contrast to the suppressed conductance at low bias for 1 M KCl, the differential conductance $dI/dV$ as a function of applied bias of 0.33 M $AlCl_3$ displays two peaks (0.216 nS at $V$=-700 mV and 0.314 nS at $V$=700 mV). We observed strong suppression for ion valence increased to $Mg^{2+}$ (**Supplementary Figure 20**) and $Al^{3+}$ shown in **Figure 3c** and **d**. For a 2 nm $MoS_2$ pore with linearized $I$-$V$ in 1 M KCl, $I$-$V$ suppression is still found for 0.33 M $AlCl_3$. This effect vanishes when further increasing the pore size to fully linearized *I-V* range where a constant conductance is observed for a 4 nm pore (**Supplementary Figure 21**). Such pore size dependent effects indicate the observation of these non-linear phenomena requires sub-nm confinements of ions.

**Multi-barrier and multi-ion conduction**

We applied classical Eyring rate model[24] to capture the strong nonlinearity in the data. Interestingly, the *I-V* fitting results in **Supplementary Figure 22** appear to suggest a multibarrier process (*n*>5) is involved in the ion transport because single barrier (*n*=1) clearly fails to find the current trends. The obtained multibarrier points out that non-mean field interactions need to be considered for this case.

We suggested here a possible interpretation for qualitatively understanding the experimental results using ionic Coulomb blockade. In the proposed transport configuration given in **Figure 1**, the free energy part can be simply written as a sum of dehydration energy $E_{dehydration}$ and electrostatic energy,

$$F_K = E_{dehydration} + E_{qQ} + E_{qq} \qquad \textbf{(Equation 1)}$$

where $E_{qQ}$ is the electrostatic interaction between the K$^+$ ion charged in *q* (+*e*, *e* is the elemental charge) with the fixed charges on the pore edge (*Q*), and $E_{qq}$ is the Coulomb repulsion for the K$^+$ ion with nearby K$^+$ ions. The multi-ion electrostatic part $E_{qQ}$ in **Equation 1** can be estimated as $E_{qQ} \sim -\chi \frac{2qQ}{\varepsilon}$ (**Equation 2**), where $\chi$ is a pore geometry related number and $\varepsilon$ is the solution dielectric constant, and the $E_{qq}$ can be obtained by $E_{qq} \sim \chi \frac{qq}{\varepsilon}$ (**Equation 3**). The total electrostatic energy change $\Delta E_{binding}$ due to adding one additional ion to the system then finds its minimum at $Q=q/2$, where current peak (oscillation) occurs. When multi-ion system is introduced, periodic energy minimums then appear at $Q=Nq/2$, depending on the total number of ions $N$ in the pore. Under this multi-ion conduction consideration, the oscillation peaks observed in the experiments may occur when a transition from $N$ ion system to $N+1$ ion system happens in the pore due to barrier-less conduction at discrete surface charge values[10]. The ionic Coulomb blockade effect is very peculiar because ions are considered as classical particles here, but the electrostatic exclusion principle for ions in sub-nm pores leads to Fermi-Dirac analogies[25].

This experimental finding recovered comparable oscillatory results predicted in both simulations and analytical theory[10,11]. As discussed in the theoretical counterparts, these peaks are out of the reach by continuum-based PNP approaches and therefore may be accounted as potential evidences for ionic Coulomb blockade oscillations. In agreement with previous predictions[10], the large oscillatory ratio obtained for multivalent ions (**Figure 2j-l**) seems to suggest a transition to strong Coulomb blockade regime due to valence amplified cation-cation Coulomb interactions. The energy difference $\Delta E$ for adjacent $I_{max}$ and $I_{min}$ current peaks can be estimated according to ionic Coulomb blockade theory approximation[12],

$$\frac{I_{min}}{I_{max}} \approx \cosh^{-2}(\frac{\Delta E}{2.5\ k_B T}) \qquad \textbf{(Equation 4)}$$

**Equation 4** is plotted in **Supplementary Figure 23**. Results in **Figure 2j-l** thus lead to ~2 to 3 $k_B T$ difference in the current fluctuations. We found the oscillations (**Figure 2**, **Supplementary Figure 11**, **Supplementary Figures 18-19**, $I_{min}/I_{max}$ from 31% to 80%) and energy difference from the experiments are still modest compared with what theory predicts or our straightforward estimations[13]. The deviation may come from short channel effect for our $MoS_2$ pore, surface effect, particle effect of surrounding water molecules, membrane charging and other nonequilibrium local conditions in the experimental system that influences the current.

**Figure 2** does not reveal a clear difference on the period of oscillations. The $MoS_2$ pore may not maintain the same surface charges after changing the solution from KCl to $AlCl_3$. Moreover, we would like to note the possibility of ion pairing also complicates the comparison between monovalent ions and trivalent ions in practical systems[11,26]. It has been proposed that the fractionalization of multivalent ions may lead to self-energy barrier comparable to that of monovalent ions[26]. This fractionalization phenomenon of trivalent ions was also observed in our simulations as shown in **Supplementary Figure 24** where one aluminum ion may bind either one or two additional chloride ions.

**All-atom molecular dynamics simulations**

To provide a further molecular level understanding of the underlying mechanism of the multi-ion Coulomb blockade in atomically thin sub-nm pores, we then developed a molecular dynamics model to elucidate the proposed transport configurations and capture the oscillatory behaviors. Our simulations (**Figure 4**, **Supplementary Figure 25-28**) mimic our experimental conditions of sub-nm pores and ionic environment, and details can be found in the **Supplementary Section vi**. Molecular dynamics simulations with modern force field typically introduce fixed partial charges on surface atoms for building the electrostatics[27,28]. As we have already determined the surface charge of MoS$_2$ pore experimentally[5], and as we can actively tune them with optoelectronic control, here we followed the experimental means to directly set up the surface charges on the active pore edge atoms (*Q*) to calculate the potential of mean force (PMF) and other energetics as shown in **Figure 4**. The simulation results indicate the energetics of ion translocation through the pore is highly surface charge dependent. Although single layer MoS$_2$ pore is only 0.65 nm thin theoretically, molecular dynamics simulations show charged atomic sites in the layer can contain several K$^+$ ions, as shown in a snapshot in **Figure 4b** where three K$^+$ ions are trapped in the pore. Transition from three K$^+$ ions to two K$^+$ ions housed inside the pore was observed in the simulations (**Figure 4b**). Setting higher surface charge values (3.5 e), the simulations display multi-ion states more frequently. The free energy profiles from PMF calculations (**Figure 4c** and **d**) reveal multiple barriers for potassium ion crossing event, suggesting that dehydration and Coulomb interactions both contribute to the process[13,29-31]. In the absence of many-body Coulomb interaction of ions, single ion drift-diffusion scenario (**Figure 1a**) gives only a single high barrier ($\Delta E_{dehydration}$) due to dehydration–the stripping of water layers near the translocating ion in the sub-nm pore (**Supplementary Movie 1**). However, we found through molecular dynamics simulations, when the charge is introduced, the central pore region acts as several potential wells, as highlighted in the free energy landscape. **Figure 4c** and **d** reveal two potential wells with depths of 4.41 $k_BT$ near the pore edges and a free energy hollow of 1.22 $k_BT$ at the pore center. These binding sites are reasonable for trapping several potassium ions in a MoS$_2$ pore. The multibarrier PMF is also in good agreement with our *Eyring* rate model fitting results given in **Supplementary Figure 22**. Similarly, we have recovered comparable simulation results obtained in the graphene cases[31-33]. The free energy profile signalises that the Coulomb interaction causes barriers via electrostatically coupling the second incoming potassium ion to the first trapped one (**Figure 4c,** and **Supplementary Movie 2**). As presented in the **Supplementary Movies 2-3**, we observed co-translocation of two ions and multi-ion conduction in the simulations. The number of ions housed in the pore and multi-ion conduction depend strongly on the charge values at the pore edge. Due to the presence of ions in the pore, an electrostatic potential barrier is formed at each side of the binding site on the pore[26,30]. Depending on the charges on the pore and its atomic configuration near the edge, multiple ions can be bound in the pore and at the pore entry. This interaction repels the incoming potassium ions external to the pore via Coulombic repulsion, providing basis for ionic Coulomb blockade.

Remarkably, the computed results for ion transport under an electric field in **Figure 4e** capture the nonlinear *I-V* characteristics[13], as a large voltage is required to overcome the charging barrier. To examine whether the proposed picture manifests the monotonic conductance phenomenon[10,11], we conducted the simulation of ionic currents for different values of *Q* (0-3.5 e). As shown in **Figure 4f**, under a field of 400 mV/nm, the simulated current increases to a peak value of 2.6 nA and 1.9 nA at -0.5 e and -2.0 e, respectively. Despite of not seeing well-defined oscillation, this rollover behaviour in molecular dynamics is comparable with Coulomb blockade predictions and the surface charge determines the transport transitions. To demonstrate that the interactions are not due to the particular setup, we modelled the $MoS_2$ pores with other shapes as well as sub-nm graphene pores. The detailed results are discussed in **Supplementary Figure 27** and **Supplementary Figure 28**, which further indicate that the current phenomenon occurs generally in atomically thin sub-nm pores regardless of pore materials and pore shapes. Ionic Coulomb blockade scenario may can explain several electrostatic forces governing ion transport behaviors that were already suggested in early enlightening works[34,35].

It is worth noting that there are still unknown properties of this system although our multi-ion Coulomb blockade discussion and molecular dynamics simulation have successfully explained our experiments from several aspects, due to the limitations in both experiments as well as theory. Current molecular dynamics does not fully capture the experimental system for example chemical movements, and the simulated results only qualitatively match the experiments. We thus discussed here also other possibilities that may contribute to the system. Ionic Coulomb blockade can be alternatively interpreted as the fractional Wien effect[11], which also manifests both the nonlinear *I-V* and conductance oscillation phenomena. We examined the formation of ion pairs in our molecular dynamics simulations but did not observe considerable ion pairing events for monovalent ions, which we attribute to the high electric field and ion selectivity in the atomically thin $MoS_2$ pores. This possibility may get pronounced for trivalent ions and weak electrolytes in long nanotubes. Other possibilities like unknown chemical substitutions, specific absorption-desorption processes or thermal drifting effect on the pore could also lead to current fluctuations. Nevertheless, such effects are too specific to be correlated with the present observations, because different individual sub-nm $MoS_2$ pores resulting from our current fabrication capabilities with potentially a variety of terminated edge chemistry, shape and charge[18] have shown such oscillations and the oscillations are reproducible for same devices in repeatable modulations. We therefore believe these alternative scenarios are unlikely to be the main cause of the current findings. Of course, precise determinations of experimental systems and more accurate theories are still needed, which we leave to future studies.

**Conclusion**

To summarize, ionic conduction oscillations have been experimentally observed in the sub-nm $MoS_2$ pores under the optoelectronic charge control. We managed to actively gate the ion transport in an artificial nanofluidic system using external stimulation, presenting important steps towards realization of advanced ionic machineries. Such experiments provide a benchmark for the examination of ion transport in atomically thin pores. Thanks to the molecular dynamics simulations, we are able to unveil the potential physical image of ionic conduction oscillations at an atomic level, which may disclose a further understanding of ionic Coulomb blockade phenomenology for sub-continuum fluidic transport physics as well as for illuminating the physics of biological ion channels.

**Author Contributions**

F.C and Z.G contributed equally to the work. F.C fabricated the devices, performed experiments and analyzed data. C.Z prepared and characterized the MoS$_2$ samples, and fabricated devices. Z.G, H.Z and R.Z designed the MD simulations. Z.G performed the MD simulations. Y.C built the laser heat transfer FEM model. X.J carried out pressure measurement. Y.L participated setup implementation and analyzed data. J.F conceived and supervised the project. J.F, R.Z, F.C, Z.G, Y.C wrote the paper. All authors contributed to the general discussions.

**Acknowledgments**

This work was financially supported by the National Natural Science Foundation of China under grant number 21974123, 11574224 and U1967217, Natural Science Foundation of Zhejiang Province under grant number LR20B050002, Fundamental Research Funds for the Central Universities (2019XZZX003-01), and Hundreds Program of Zhejiang University. R.Z also acknowledges the financial support from W.M. Keck Foundation (Grant award 2019-2022) and the IBM BlueGene Science Program (Grants W125859, W1464125 and W1464164). We thank the Microfabrication Platform and Analytical Service Center at Zhejiang University for facility support.

**Data Statement**

The data that support the findings of this study are included in the manuscript, supplementary materials and are available from the corresponding author upon reasonable request.

# FIGURES

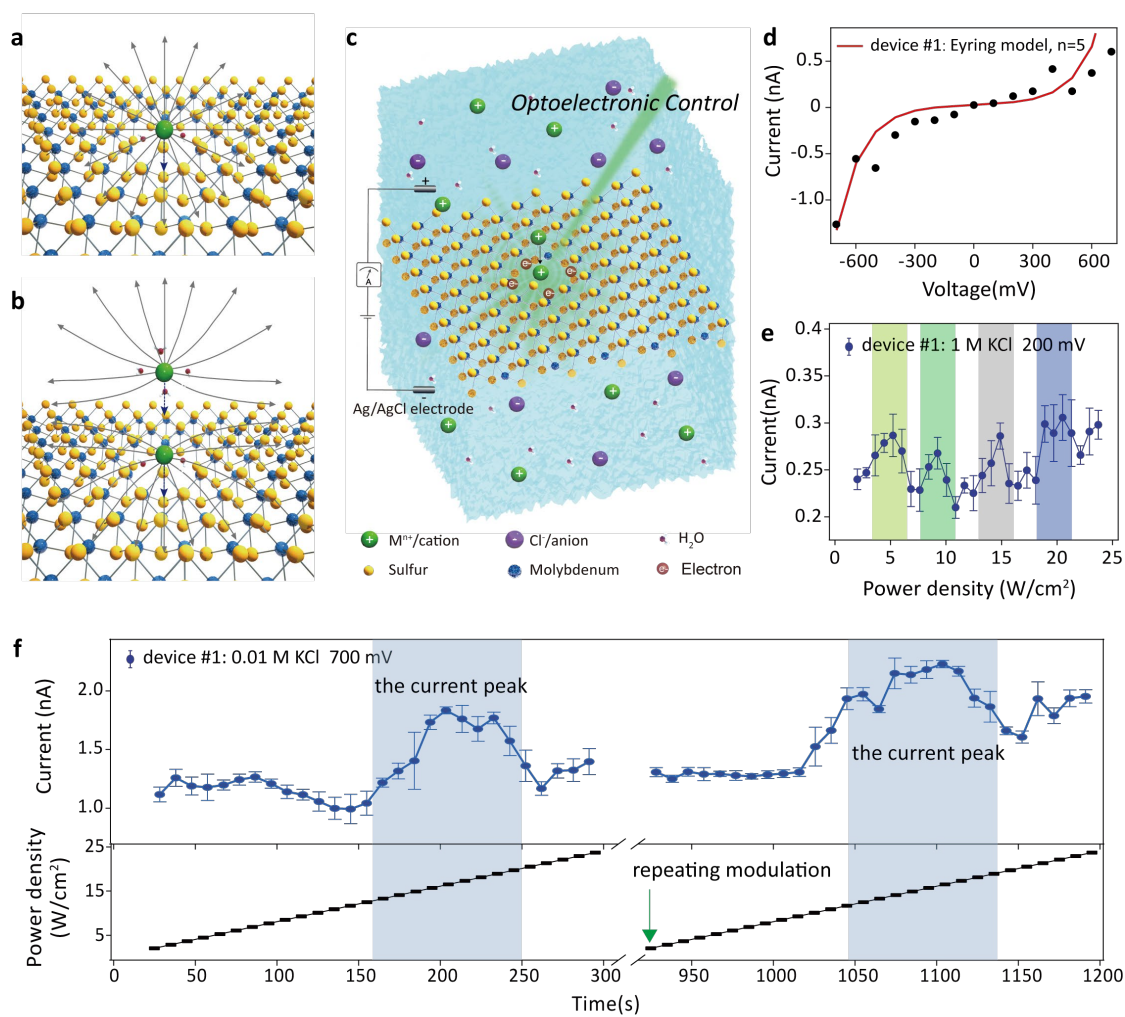

**Figure 1. Optoelectronic modulation of single ion transport in MoS$_2$ pores. a.** Single ion transport. **b.** Multi-ion conduction where ion-ion interaction is introduced. **c.** Modulation of ionic transport through sub-nm MoS$_2$ pores using optoelectronic control of surface charges. **d.** A nonlinear *I-V* characteristic of a sub-nm MoS$_2$ pore in 1 M potassium chloride aqueous solution (KCl). **e.** Periodic ionic current oscillations as a function of light power density, recorded at a fixed bias of 200 mV in 1 M KCl. **f.** Repeated current oscillations as a function of time and applied light power density, recorded at fixed bias of 700 mV in 0.01 M KCl. During the second cycle, the current oscillation peak recovers at power density of 16.5 W/cm$^2$. The current peaks are marked with color shadow. The data is taken from device #1 (<0.6 nm).

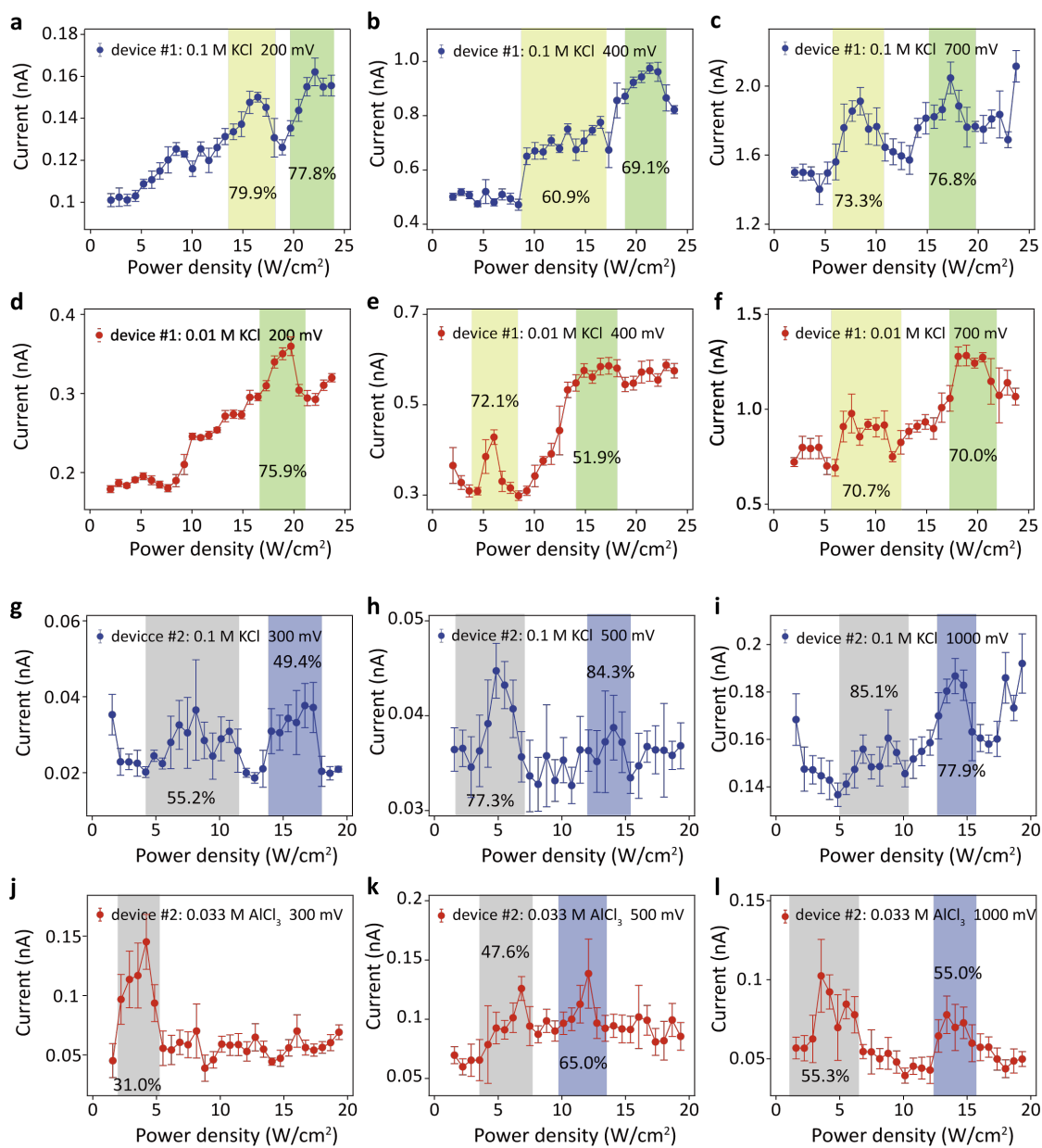

**Figure 2. Concentration and valence dependent ionic conduction oscillations. a-f.** Comparison of ionic current oscillations as a function of light power density in sub-nm MoS$_2$ pores (device #1: < 0.6 nm) in 0.1 M KCl and 0.01 M KCl recorded at 200 mV, 400 mV, 700 mV, respectively. **g-l.** Comparison of ionic current oscillations (device #2: < 0.6 nm) in 0.1 M KCl and in 0.033 M AlCl$_3$ recorded at 300 mV, 500 mV, 1000 mV, respectively. The current peaks are marked with color shadows, used for calculating oscillation ratio ($I_{min}/I_{max}$).

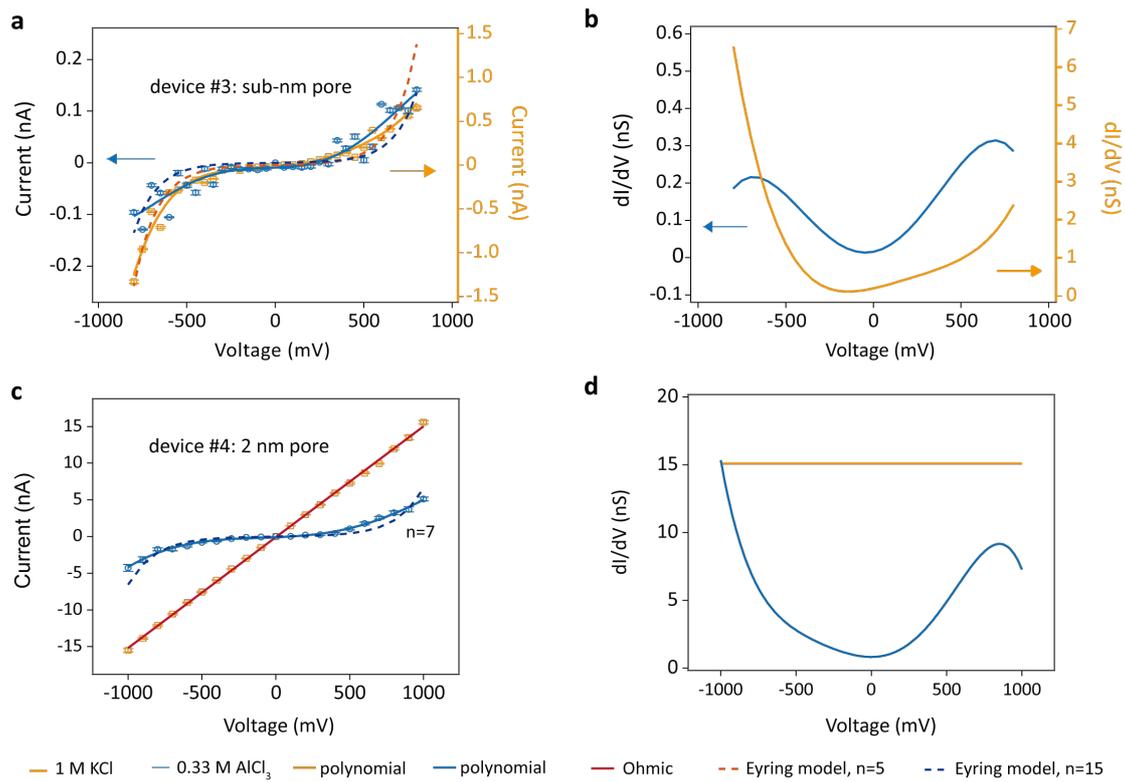

**Figure 3. *I-V*s comparison and differential conductance *dI/dV vs V*. a.** Nonlinear IV characteristics of a sub-nm MoS$_2$ pore (device #3, < 0.6 nm) in aqueous solutions under ionic strength of 1 M KCl, and 0.33 M AlCl$_3$ at room temperature conditions. **b.** The corresponding differential conductance (from best polynomial fit) *dI/dV vs V* of 1 M KCl and 0.33 M AlCl$_3$ for data shown in **a**. *dI/dV* turnover is found for AlCl$_3$ at *V*=700 mV, and *V*=-700 mV. **c.** *I-V* characteristics of a 2 nm MoS$_2$ pore (device #4, 2 nm) under in 1 M KCl, and 0.33 M AlCl$_3$. **d.** The corresponding differential conductance *dI/dV vs V* plot for the 2 nm MoS$_2$ pore *I-V* data shown in **c**.

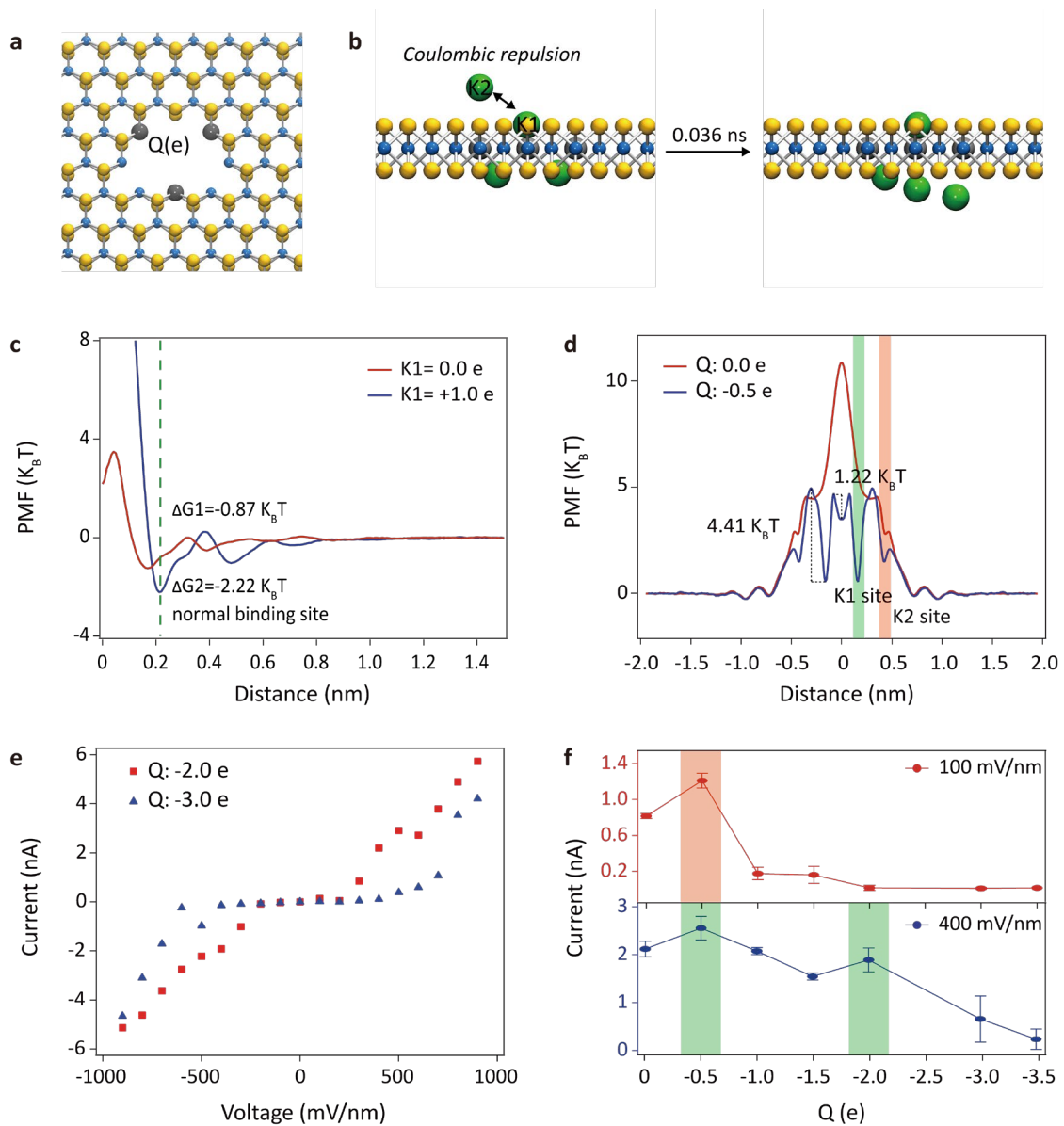

**Figure 4. Molecular dynamics simulations of ionic transport through MoS$_2$ pores. a.** Snapshot of a sub-nm MoS$_2$ pore with charged edge atoms (marked by grey atoms ***Q***). **b.** Schematic of potassium ions passing through MoS$_2$ nanopore with ion-ion interactions and the corresponding dynamic process can be found in the **Supplementary movie 2**. The K+ ions are shown with turquoise spheres, wherein the key two ones are defined as K1 and K2. **c.** The PMFs of K1 site (0 e and +1.0 e) trapped in the nanopore. Specifically, the PMFs evaluate the free energy alterations by fixing K1 ion while pulling K2 ion along the direction perpendicular to MoS$_2$, according to the positions of two ions in the first picture of **Figure 4 b** − K2 ion is pulled along the z-axis. The reaction coordinate (distance) indicates the distance between K1 and K2 ions along the direction perpendicular to MoS$_2$ pore. **d.** Free energy profiles of a K+ ion

passing through sub-nm pores with $Q$ (0 e and -0.5 e). The reaction coordinate (distance) indicates the vertical interval of the K$^+$ to the pore center. The green and orange shadows mark the K1 and K2 sites. **e.** Simulated nonlinear *I-V* characteristics of MoS$_2$ pores with $Q$ of -2.0 e and -3.0 e. **f.** Representative curves of ionic current *vs* pore charge under 100 mV/nm and 400 mV/nm electric field. Current peaks from the simulations are marked with color shadow.

# Supplementary Information (SI) for:
# Ionic conductance oscillations in sub-nanometer pores probed by optoelectronic control

Fanfan Chen[1#], Zonglin Gu[2#], Chunxiao Zhao[1], Yuang Chen[1], Xiaowei Jiang[1], Zhi He[2], Yuxian Lu[1], Ruhong Zhou[2,3] and Jiandong Feng[1]*

[1]Laboratory of Experimental Physical Biology, Department of Chemistry, Zhejiang University, 310027 Hangzhou, China

[2]Institute of Quantitative Biology, College of Life Sciences and Department of Physics, Zhejiang University, 310027 Hangzhou, China

Department of Chemistry, Colombia University, New York, NY10027, USA

*Correspondence should be addressed to jiandong.feng@zju.edu.cn

**Table of contents:**

i. Methods

The sub-nanometer MoS$_2$ pores were fabricated and characterized using techniques as reported elsewhere[1]. More detailed information can be found in the published protocol[2]. Briefly, CVD grown single layer MoS$_2$ membranes are transferred to substrates with a predesigned 60 nm supporting hole in 20 nm thick silicon nitride membrane (Norcarda Chips). The ECR approach is then employed to open pores in an atom by atom manner[1-4]. In this work, we succeeded in making roughly hundred sub-nm MoS$_2$ pore devices. The ion transport measurements were performed using an Axopatch 200B patch-clamp amplifier (Molecular Devices) and PXI data acquisition card (National Instruments) (recording frequency 6,250 Hz) , as described previously[1,4]. The MoS$_2$ nanopore chip is mounted in a PMMA flow cell and exposed to light illumination from a 488/532/785 nm solid state laser (OBIS Coherent and Cobolt, measured long-term power stability with std < 1.6 %). The flow cell is aligned to the laser beam under a USB microscope which is placed opposite to the flow cell. We achieved control of MoS$_2$ pore surface charges by modulating the laser power density using an external signal generator (analog modulation), synchronized together with ionic current recordings. The beam position is aligned to the center of the chip with a micromanipulator, where the power density is measured from an optical beam profiler. All measurements were performed under room temperature. Pore sizes were estimated by substituting conductance into the established formula[2,5] and compared with the conductance of 1 nm pore calibrated in 1M KCl. Due to scattering and refraction in the flow cell, the estimated light power arriving on the chip surface is obtained by *P=0.935\*P$_0$* due to transmission loss through the fluidic cell, where *P$_0$* is the original laser power. For light power density calculation, we use two-dimensional Gaussian orthographic fitting to estimate the power density that reaches the surface of the membrane. The two-dimensional Gaussian orthographic function is,

$$f(PSF) = A \times exp\left[-\frac{1}{2}(\frac{(x-\mu_1)^2}{\sigma_1^2} + \frac{(y-\mu_2)^2}{\sigma_2^2})\right] + B$$

where the $\sigma_1, \sigma_2$ is the full width at half maximum (FWHM); A is the peak value of laser power; $\mu_1, \mu_2$ is the location of peak value; and $B$ is the background. The Gaussian beam is considered in the power density estimation. The detailed discussion is shown in **Supplementary Figure 3** and its captions. It shall be noted a direct division of power by the total area without considering the Gaussian beam may underestimate the real power density. A supplementary movie is included to show the laser alignment process.

*I-V* measurements are performed before and after the experiments for checking the pore size. Detailed conditions for each figure are mentioned in the captions. pH of all solutions stays unbuffered (1 M KCl dissolved in DI water, pH about 5.5) to avoid introducing other molecular species. All data analysis about ionic current oscillations as a function of light power density (recording conditions are marked in the Figures or Captions) have been done using custom-made Matlab (R2016a) code. The nonlinear *I-V*s are fitted based on Eyring rate model[6] and number of barriers is directly obtained from fitting. *dI/dV* plot is made from the best polynomial fit. The pore diameter is estimated from the conductance with an error of $\pm 0.2$ nm[7]. The pore size and device number information has been added to relevant figure. For pores smaller than 0.6 nm, we referred directly as pores < 0.6 nm due to both the error and the limitation of characterization capabilities. Error bar shown in the ionic current data is obtained based on analysis of the standard deviation of sampled current trace in 10 s or 20 s recordings.

ii. Numerical Simulation on Laser Induced Heating

COMSOL Multiphysics 5.2 was used for the finite element method (FEM) simulation. We used the *heat transfer in solids* as physics to calculate a stationary solution for coupled heat transfer of solids and fluids. **Supplementary Figure 16** shows the geometry in the simulation. The silicon chip was encompassed by a 10 mm cubic PMMA flow-cell and defined as 280 μm thick and 5×5 mm square. The backside of silicon chip contained a 585 μm large trapezoid opening penetrating through the chip, leaving a 12 μm large membrane area on the front side. Liquid chambers for electrodes are 4 mm apart from the chip in the x direction and all chambers are 1.5 mm in diameter. In the *heat transfer in solids* module, initial temperatures were set to 298.15 K. *Heat Transfer in Solids* and *Heat Transfer in Fluids* were used along the silicon chip, the PMMA and the liquid. The laser-induced heating was realized by using two *Deposited Beam Power* modules. First one was applied to the area outside of the 12 μm large membrane with power of 70% of the laser power (silicon absorbance), while the second one was applied to the membrane only with power of 10% of the laser power (silicon nitride absorption)[8]. Both laser beams are Gaussian distributed and centered in the middle of the membrane with a standard deviation 200 μm. Furthermore, since the silicon chip have a 60 nm thick silicon dioxide layer and a 20 nm thick silicon nitride (outermost layer), reflection of different media surfaces and quantum yield of $SiN_x$ film were also taken into account. Therefore the energy flux of the laser was calculated using $P = P_0 \times A \times (1 - q) \times (1 - R)$, where $P_0$ was the power of the incident laser, $A$ was the absorbance, $q$ was the quantum yield of $SiN_x$ film (0.07) and $R$ was the reflection[9]. Reflection was calculated using $R = (n_1 - n_2)^2/(n_1 + n_2)^2$, where $n_1$ and $n_2$ are refractive index of different media, and multiple reflections are multiplicative. As a boundary condition for the temperature, all surfaces of the PMMA flow-cell were set to be 298.15 K. Then, the finite element simulation was run for different $P_0$ by parametric sweep.

In the FEM simulation, parameters in Deposited Beam Power modules and determination of boundary condition of FEM equation are dominant factors that influence the increase of temperature induced by laser-heating effect. Different from previous reports, we used a relatively large standard deviation of 200 μm laser beam instead of a focused laser beam, which changed the energy flux of laser $P$ in Deposited Beam Power modules by changing $A$ (absorbance) greatly as the equation shown above. This change is because there is an obvious difference of absorbance between the silicon (70%) and the silicon nitride membrane (10%), thus the power of a large beam laser (in our experiment) will be absorbed largely but the focused laser beam will concentrate its energy on the low absorbance silicon nitride membrane, which is more like the condition in optical traps, where the heating effect is rather small when using laser powers of less than 100 mW. Therefore, temperature increase in our FEM simulation is about an order of magnitude higher than the experimental results with a focused laser or in optical traps. Even so, it is worth noting that the temperature increase of about 4.5 K (at peak value when $P_0 = 19.72$ W/cm$^2$) is the highest possible temperature increase since we used the most conservative temperature boundary condition estimate (all surfaces of PMMA flow-cell were set to be 298.15 K), far away to induce current difference given in the data. Moreover, since the effect of temperature increase on the ionic conductivity is monotonic, increase in temperature may not cause the phenomenon of current oscillations. Thus, consistent with several other reports[8,10,11], we conclude the laser induced heating is a minor effect for our nanopore system.

iii. Surface Charge Modulation

To know how the light affects the surface charge of MoS$_2$ pores, we characterized conductance measurement under various ionic concentrations. Analysis of surface charge density was estimated by referring to the model built by Lee et al.[12],

$$G = \kappa_b \left[ \frac{4L}{\pi d^2} \times \frac{1}{1+4\frac{l_{Du}}{d}} + \frac{2}{\alpha d + \beta l_{Du}} \right]^{-1},$$

where $\kappa_b$ is the bulk conductivity; $L$ is the pore length and $d$ is the pore diameter; $\alpha = \beta = 2$ in the calculation to get the best fitting; $l_{Du}$ is the Dukhin length. From the fitting result shown in **Supplementary Figure 8**, the surface charge density is modulated from 15.4 to 37.4 mC/m$^2$ by laser illumination, in agreement with reported values[8].

Reversal/osmotic potential measurement also provides another clear evidence for the effective modulation of surface charges resulting from the ion selectivity of MoS$_2$ nanopore[8]. The MoS$_2$ pore is known to be negatively charged[4,8], leading to efficient cation ion selectivity. With introducing the 532 nm laser illumination, **Supplementary Figure 9** shows the osmotic potential changes from -59.1 mV to -92.2 mV, suggesting the increase of negative charges on the surface[8]. The surface charge density is proportional to the Dukhin length according to the function of $l_{Du} = \frac{\sigma}{2 \times c_s \times e}$, suggesting the increasing surface charge under illumination, where the $\sigma$ is the surface charge density; $c_s$ is the bulk ion concentration; $e$ is the elementary charge.

Light MoS$_2$ interaction is further identified by photoluminescence measurements (**Supplementary Figure 29**). We used a 532 nm laser (LabRAM HR Evolution) to excite CVD prepared MoS$_2$ samples on SiO$_2$/Si substrate, and the strongest photoluminescence emission is found for monolayer MoS$_2$ samples that we use for fabricating the pore devices (**Supplementary Figure 1**).

## iv. Oscillating Current Modulation vs. Random Fluctuations

*Reproducible patterns*: First, the phenomenon of oscillating current patterns is observed in sub-nm MoS$_2$ pores. **Supplementary Figure 4** shows the light induce the ionic current change is obvious compared to the baseline noise level. Moreover, we record each light modulated current experiments for at least two independent measurement cycles. The reproduced oscillating current pattern shapes with continuous cycles in **Figure 1** of main manuscript and **Supplementary Figure 11** emphasized again the recovered patterns are due to active modulation from light control.

*Pore growth*: There is another possibility that laser might damage the MoS$_2$ sample and enlarge the pore. We carefully select the power range and the illumination beam condition to avoid this pore growth, which is in good agreement with previous work[8]. During the measurements, when power density is reset to initial value the current gets recovered to nearly the same starting values as show in these data (**Figure 1, Supplementary Figures 11, 12-15**), which further suggests the pore does not grow up.

*Noise spectral analysis*: We also compared the noise spectra of the sub-nm pores under various conditions as the low frequency noise (flicker noise/pink noise) could originate from surface charge fluctuation. The dependency could be described as the Hooge's formula[13],

$$s(f) = \frac{A_H}{f} = \frac{\alpha \bar{I}^2}{f}$$

where A$_H$ is the low frequency noise amplitude parameter[14]; $\bar{I}$ is the mean ionic current; $\alpha$ is the noise amplitude. The fitting result is presented in **Supplementary Figure 19**.

*Nanobubbles, wetting and other issues*: The noise spectral analysis in **Supplementary Figures 5-7** show the noise in our devices is normal where power density spectra show 1-10 pA$^2$/ Hz in dark conditions at low frequency side, which agrees well with reported noise levels for MoS$_2$ pores suspended on a 60/70 nm SiN$_x$ supporting hole[2]. (Supporting hole size also significantly influences *1/f* noise). Usually, poor wetting or nanobubbles dramatically introduces high noise in the low frequency range (PSD increases several orders of magnitudes)[15,16], which does not appear in the shown noise spectra. Moreover, we applied pressure regulation to probe the possibility of nanobubble or bad wetting in our sub-nm pores. This result is included in **Supplementary Figure 7**. The application of pressure does not cause the low frequency noise change of the system (fitting shown in **Supplementary Figure 7c**) which we attribute to good wetting for the current case. Finally, the bubble or wetting related issues are stochastic, lacking the reversibility and control in time. The actively controlled oscillation patterns also suggest the current results may not origin from such situations.

## v. Fitting of Non-linear IV Data

The nonlinear IV data shown for the sub-nm pores can be fitted using a modified Eyring rate theory,

$$I = K \frac{\exp\left(\frac{zF}{RT}\frac{V-a}{n}-1\right)}{\exp\left[\frac{zF}{RT}(V-a)-1\right]} \left[\exp\left(\frac{zF}{RT}V\right)-1\right]$$

To be more specific, the parameter $n$ is the number of transition barriers[6] of the ion going through the membrane. The parameter $K = Fzk\lambda C_0$, $F$ is the faraday constant, $z$ is the ionic valence number, $k$ is the rate constant, $\lambda$ is the distances between minima of potential energy, $C_0$ is the concentration of solution. The parameter $a = -ng\frac{RT}{Fz}$, $g$ is the value for the barrier height increment, $R$ is gas constant; $T$ is the temperature.

## vi. Molecular Dynamics Simulations

In this study, three nanopore configurations were utilized to probe the ionic Coulomb blockade through all-atomic MD simulation approach. Specifically, two holes were separately embedded in two $MoS_2$ nanosheets of a same dimension (6.3 × 6.5 $nm^2$), with the trapezoid pore consisting of 7 molybdenum atoms and 2 sulfur atoms in the pore boundary (**Figure 4a**) and another hexagon pore comprising 3 molybdenum atoms and 6 sulfur atoms in the pore boundary (**Supplementary Figure 27a**). The third hole was drilled in a graphene nanosheet with a size of 6.1 × 6.4 $nm^2$ as indicated in **Supplementary Figure 28a**, which contained 9 carbon atoms in the boundary. Then, three nanopores were placed in three boxes (heights of ~ 6.0 nm) including 1 M KCl, yielding three systems separately named sys-1 (trapezoid $MoS_2$ pore system), sys-2 (hexagon $MoS_2$ pore system) and sys-3 (graphene pore system). To detect the effect of nanopore charges to the transport characteristics of ions as demonstrated in our experimental findings, some atoms in the pore boundary were endowed several charges Q. For sys-1, we conducted three simulations with each altering the charges (0-3.5e) of three discrete molybdenum atoms in the pore as shown in **Figure 4**. For sys-2, we performed four simulations with each altering the charges (0e, -0.5e, -1.0e and -1.5e) of three molybdenum atoms in the pore as shown in **Supplementary Figure 27**. For sys-3, we firstly carried out four simulations by changing the charges of each atom in the pore (as **Supplementary Figure 28b**) and then carried out six simulations by changing the charges of carbon 1, 4 and 7 in the pore (**Supplementary Figure 28c**). In all simulations, a series of electric fields (varying from -900 mV/nm to 900 mV/nm with an interval of 100 mV/nm) were applied along the direction vertical to the nanosheet.

The MD simulations were carried out with the software package GROMACS (version 4.6.6).[17] The VMD software was used to analyze and visualize the simulation results.[18] We adopted the CHARMM36 force field[19] and TIP3P water model[20] for the ions and water molecules, respectively. The force field parameters of $MoS_2$ and graphene were separately derived from previous studies.[21,22] The atoms of $MoS_2$ and graphene were fixed throughout all simulations. The temperature was fixed at 300 K using a v-rescale thermostat[23] and the pressure was maintained at 1 atm along the direction perpendicular to the nanosheet by using a semi-isotropic Parrinello-Rahman pressostat (x+y, z).[24] Periodic boundary conditions were implemented in all directions (x, y and z). The long-range electrostatic interactions were treated with the PME method,[25] and the van der Waals (vdW) interactions were calculated with a cutoff distance of 1.2 nm. The water geometry was constrained using the SETTLE algorithm.[26] Each simulation was performed for 10 ns with a time step of 2.0 fs, and data were collected every 2 ps. The ionic currents emerging from the transport of ion through nanopores under bias voltages were calculated based on $Q = \int I(t)\, dt$, where Q indicates the total transported charges.[27]

The PMF values of ions along the direction perpendicular to the nanosheet surface in different simulations were calculated using umbrella sampling simulations.[28-30] The distances (d) to the initial sites were restrained at a reference distance ($d_0$) with a harmonic force,

$$F = k \times (d - d_0)$$

where k was the force constant (ranging from 2000 ~ 8000 kJ mol$^{-1}$ nm$^{-2}$). To be specific, in the cases of **Figure 4c** and **Supplementary Figure 28e**, the distance indicated the vertical interval between two potassium ions (K1 and K2) along the direction perpendicular to the nanosheet; whereas in the cases of ion passing through the nanopores (i.e., **Figure 4d**, **Supplementary Figure 27c** and **Supplementary Figure 28f**), the distance denoted the vertical interval between the transport ion and the nanopore center along the direction perpendicular to the nanosheet. The spacing of the sampling windows was 0.1 nm. At each $d_0$, the system was equilibrated for 2 ns, followed by a 10 ns productive run (**Supplementary Figure 25**). The free energy profiles were obtained by the g_wham tool that implements the Weighted Histogram Analysis Method.[31-33]

# viii. Supplementary Figures

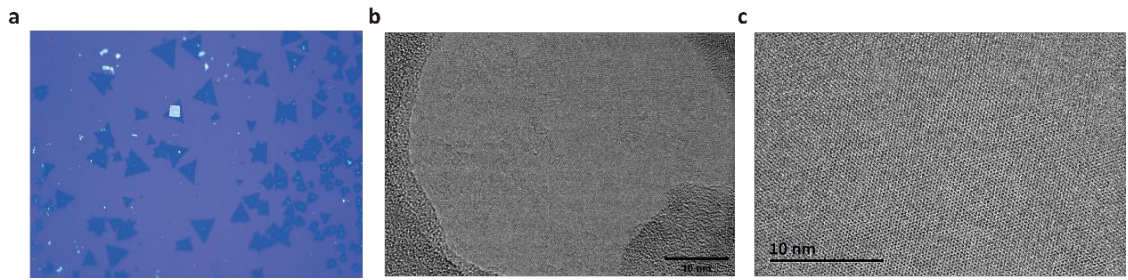

**Supplementary Figure 1. Characterization of the MoS$_2$ samples. a.** Optical image of transferred single crystal monolayer MoS$_2$ samples on silicon nitride membrane. **b** and **c.** TEM images for intact single layer MoS$_2$ membrane before applying ECR atom by atom pore opening. More details on device fabrication and characterization can be found elsewhere[1-3].

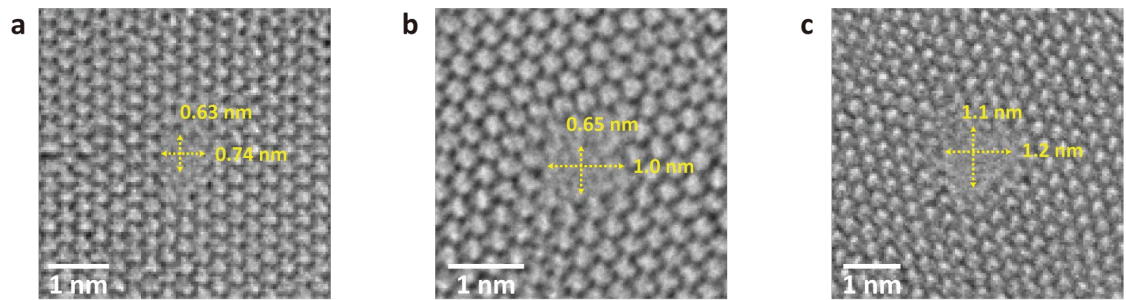

**Supplementary Figure 2. TEM image of the sub-nanopores.** Images in **a**, **b** and **c** show nanopores with diameter 0.7, 0.8 nm and 1.2 nm.

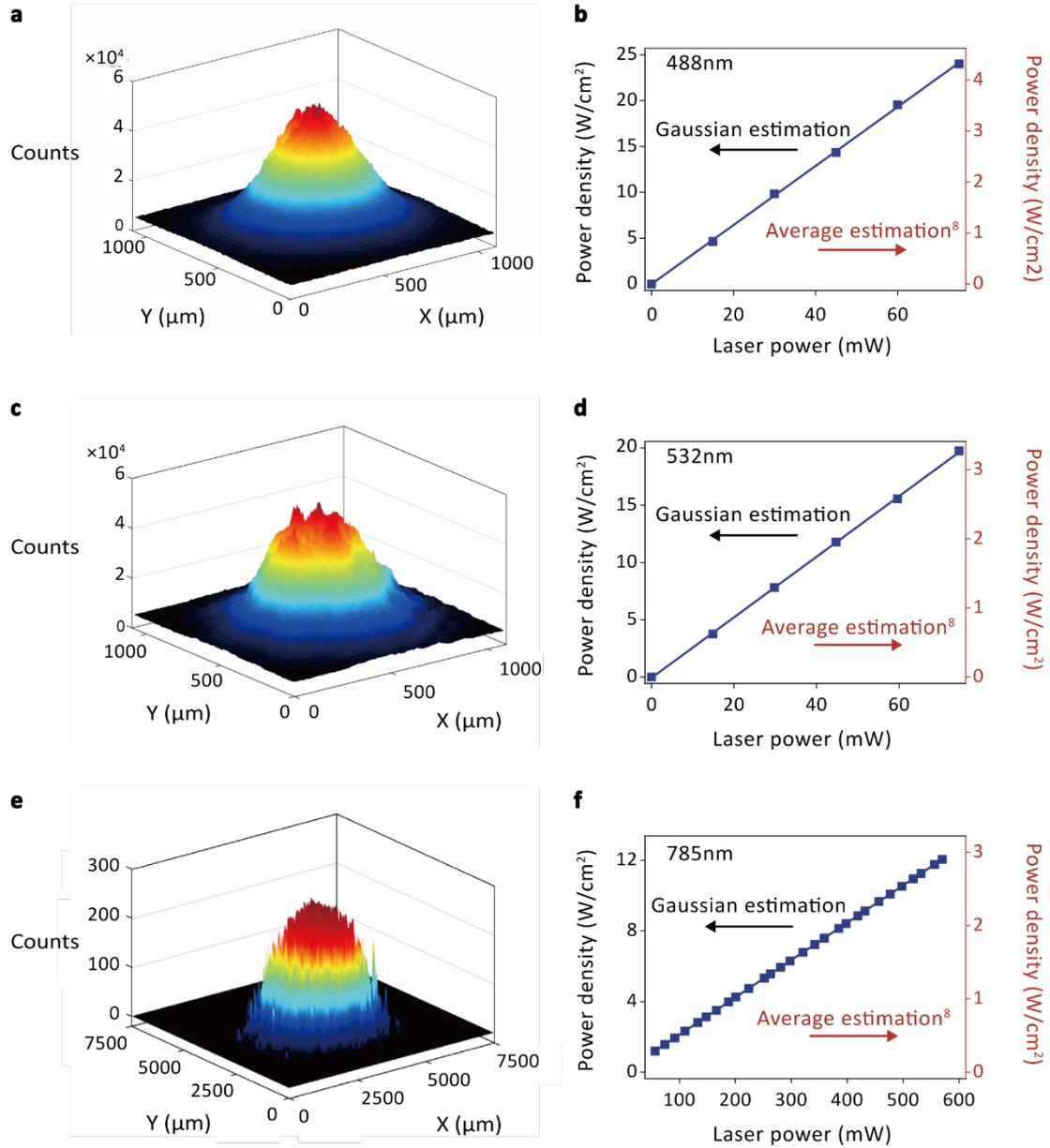

**Supplementary Figure 3. Light power density determination from the laser beam profile.**
**a.** 3D shape of our laser beam at 488 nm. **b.** light power density under analogy modulation for 488 nm. **c.** 3D shape of laser beam at 532 nm. **d.** 532 nm laser light power density under analog modulation. **e.** 3D shape of laser beam at 785 nm. **d.** 785 nm laser light power density as a function of laser power. Using Gaussian fit, we calculate the light power density as follows:

$$I = P \times \frac{\iint_{\frac{\pi d^2}{4}} f(PSF)}{\iint_{\frac{\pi D^2}{4}} f(PSF)} \times \frac{4}{\pi d^2}$$

where D is the diameter of light spot (here, D=1.1 mm); f(PSF) is the result of two-dimensional Gaussian normal distribution of light spot; d is the control accuracy of our alignment (here, d=0.22 mm). We would like to note if one uses another estimation approach where $I = P \times \frac{1}{\pi \times X_{4-\sigma} \times Y_{4-\sigma}}$, without consideration of Gaussian beam distribution, a relatively lower density number in the range of 0-3.3 W/cm² would be then obtained for the same data, compared with the range of 0-19.7 W/cm² that we estimated here at 532 nm laser. Thus, the light power density we apply is in good agreement with previous work[8], which is considered to be a safe range avoiding MoS$_2$ pore growth. We include a movie showing the alignment of laser center to the membrane in the end of this SI section.

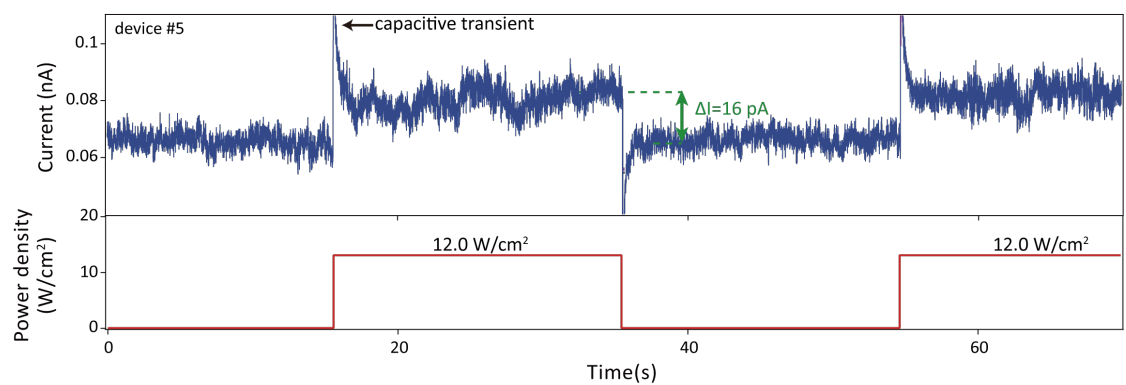

**Supplementary Figure 4. Light-induced reversible ion current modulation for a sub-nm MoS$_2$ pore.** Time trace of ionic current in 1 M KCl under modulation of 'laser on' (488 nm, 12.0 W/cm$^2$) and 'laser off', recorded at 200 mV (device #5: <0.6 nm). (The current trace was recorded at sampling rate 6250 Hz with a low pass filter set to 100 Hz)

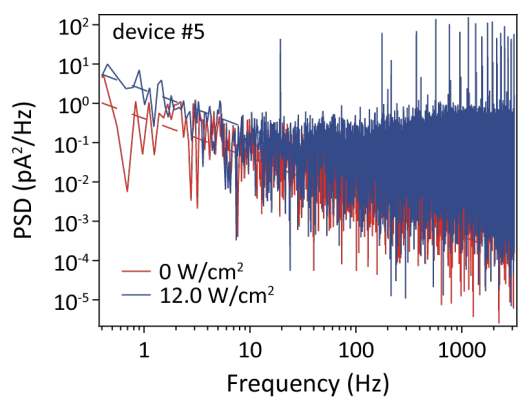

**Supplementary Figure 5. Noise spectral analysis under light modulation.** Example of current noise spectra in 1 M KCl under modulation of 'laser on' (488 nm, 12.0W/cm$^2$) and 'laser off', recorded at 200 mV (device #5: <0.6 nm). Application of light slightly increases the low frequency noise. The current trace was recorded at sampling rate 6250 Hz and the Root Mean Square (RMS) is 24.26 pA with light and 12.67 pA without light.

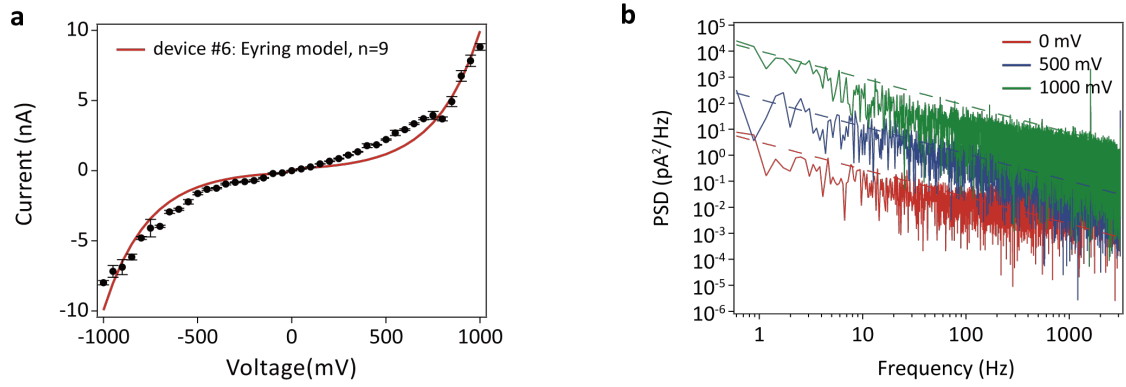

**Supplementary Figure 6. Noise analysis under different applied voltages**. **a.** Non-linear *I-V* characteristic of MoS$_2$ nanopore (device #6: 0.7 nm). **b.** Typical current noise spectra in 1 M KCl under applied voltage. The value of RMS is 19.5 pA at 0 mV, 38.1 pA at 500 mV, 152.6 pA at 1000 mV. The typical noise properties in low frequency band also suggest there is no bubble or pore wetting issues.

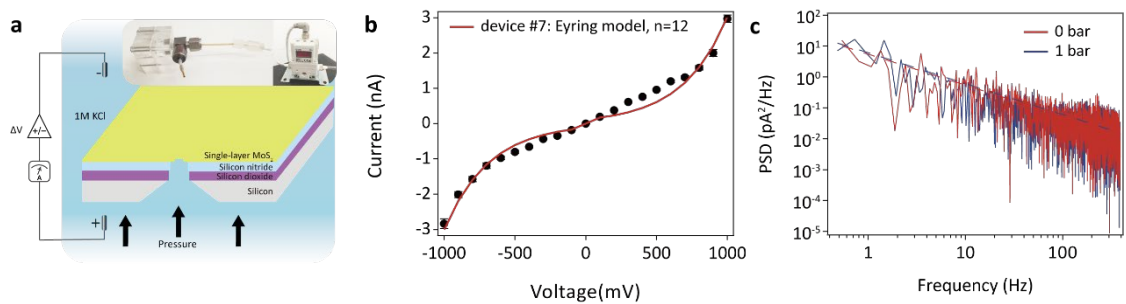

**Supplementary Figure 7. Noise analysis under pressure application**. **a.** The schematic of ionic transport through sub-nm $MoS_2$ pores regulated by application of pressure. The inset shows the experiment set-up for integration of pressure to the flow cell. **b.** Non-linear *I-V* characteristics of a sub-nm $MoS_2$ pore in 1 M KCl (device #7: <0.6 nm). **c.** Typical current noise spectra under pressure modulation in 1 M KCl recorded at 100 mV. The low frequency noises for both cases are quite similar, suggesting again no wetting issues or bubbles in the system.

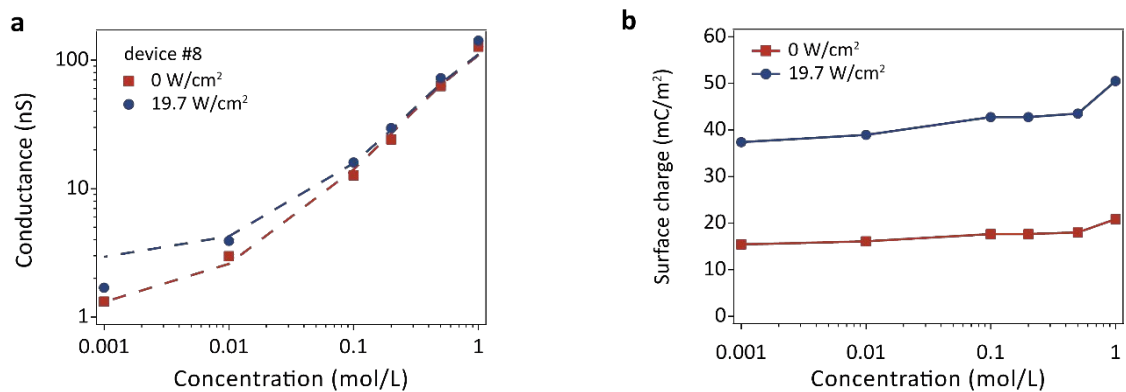

**Supplementary Figure 8. Surface charge measurements.** Conductance as a function of salt concentration with different light power density at 532 nm. And we find the extracted surface charge values to be -15.4 mC/m$^2$, -37.4mC/m$^2$ for dark and light illumination conditions for a 12.5 nm MoS$_2$ pore (device #8). Details of surface charge analysis is shown in the SI surface charge modulation discussion.

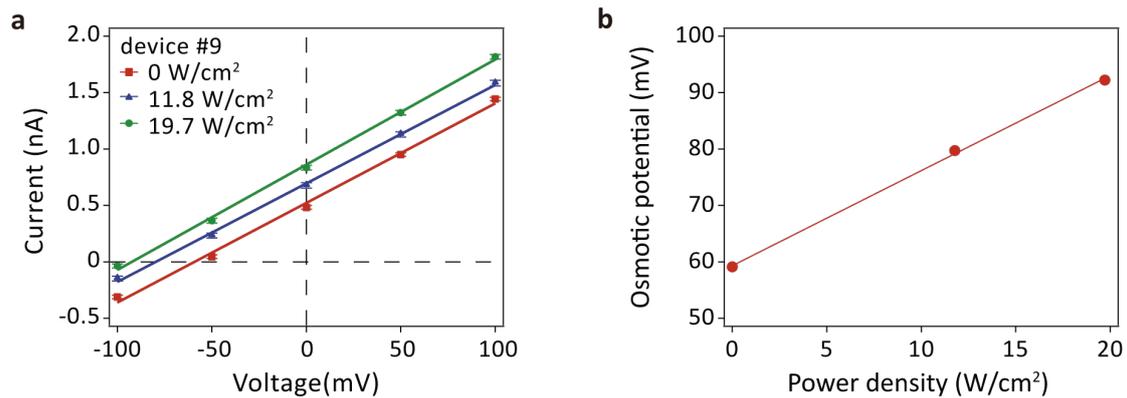

**Supplementary Figure 9. Reversal potential measurements. a.** *I-V* characteristic for a 2 nm MoS$_2$ nanopore (device #9) in a 1 M/1 mM KCl gradient under different light power density at 532 nm. b. Reversal/osmotic potential modulated by the light control.

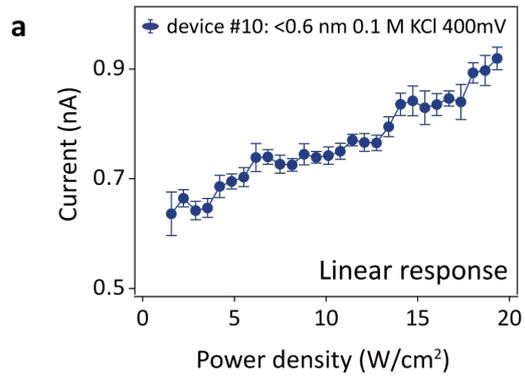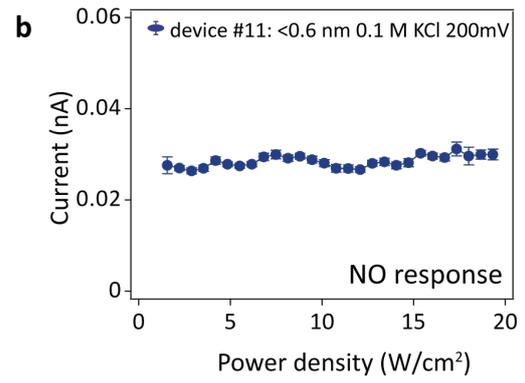

**Supplementary Figure 10. Other types of modulation responses. a.** Linear response to optical stimulation. **b.** No response to optical stimulation.

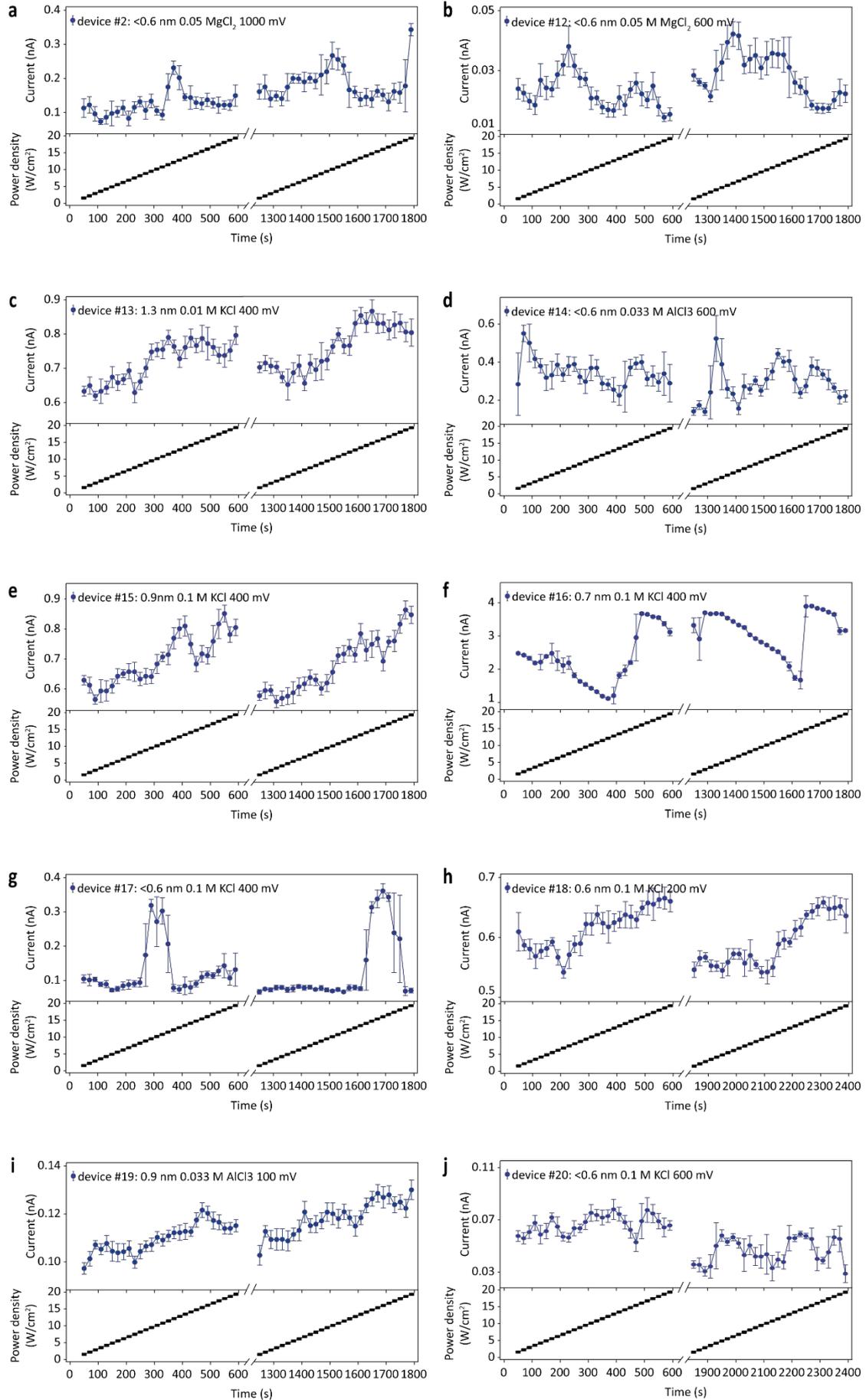

**Supplementary Figure 11. Repeated oscillatory patterns in time from 10 MoS$_2$ pore devices.** Each light modulation measurement is performed twice in a cycled modulation way in time. **a-j** are represented recordings from 10 different MoS$_2$ pore devices for recovered pattern shapes. These pore sizes are different so we did not compare the concentration and valency effect here. These results indicate the ionic conduction oscillations patterns are actively controlled by light-based surface charge modulation and the pores are not enlarged by the laser application as both pattern shapes and initial current point are recoverable.

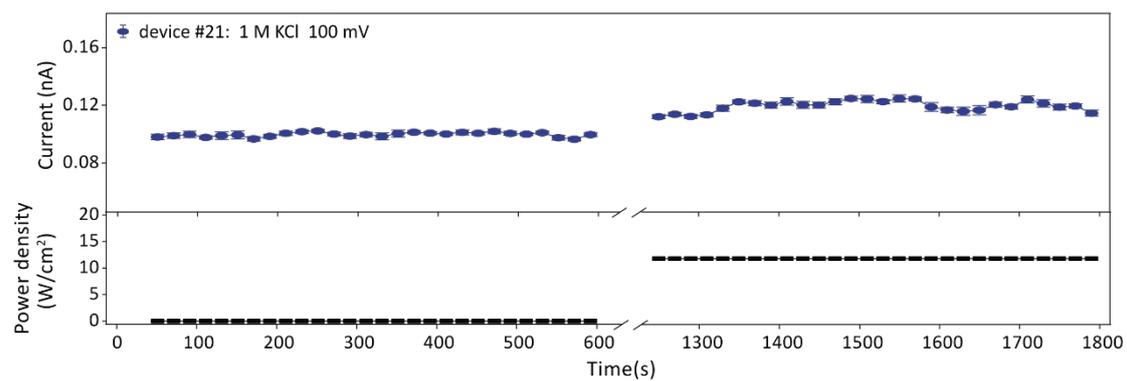

**Supplementary Figure 12. Current baseline stability in a sub-nm MoS$_2$ pore (device #21 <0.6 nm).** The noise induced variation is found to be relatively small for the both control experiments in dark and constant laser power conditions ($I_{std}$ < 3.3%).

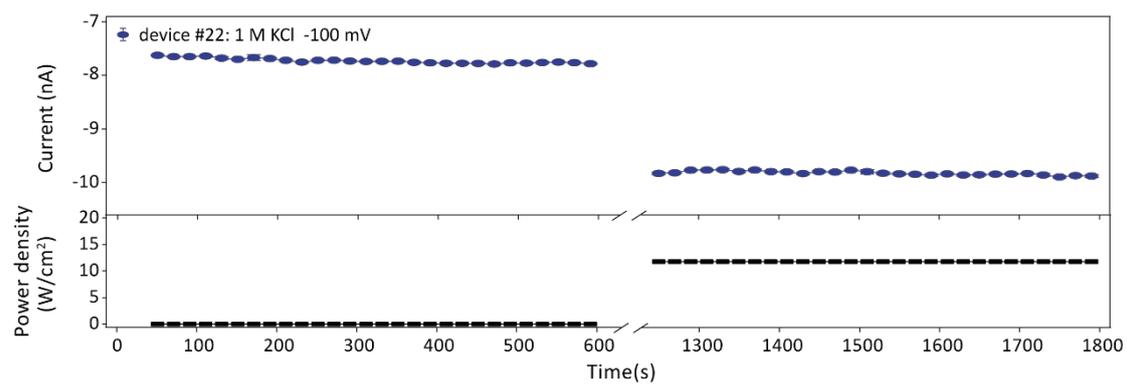

**Supplementary Figure 13. Long-term ionic current stability in large MoS$_2$ pore.** Time trace of ionic current in 1 M KCl under dark and constant laser power conditions at 532 nm recorded at -100 mV, pore size is 9 nm (device #22).

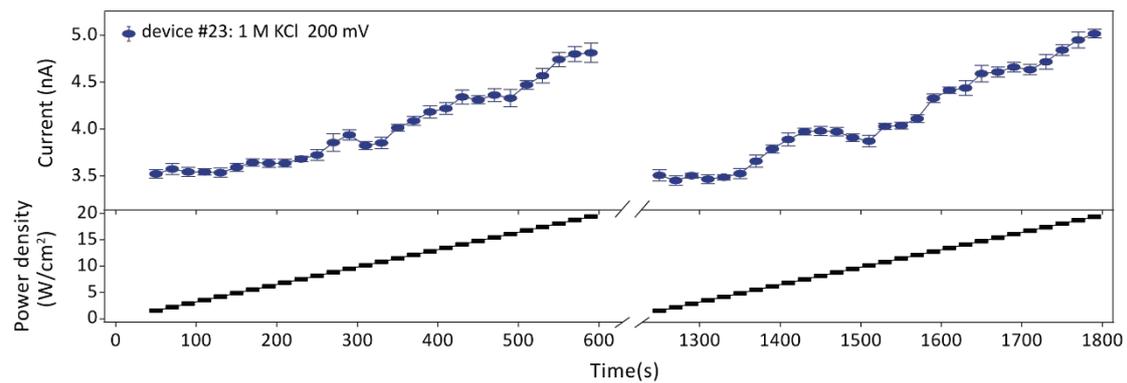

**Supplementary Figure 14. Ionic current optoelectronic modulation in a 1.4 nm MoS$_2$ pore.** Ionic current as a function of time in a 1.4 nm pore, recorded at fixed bias of 200 mV in 1 M KCl with two cycles, when light power density changes as a function of time (device #23).

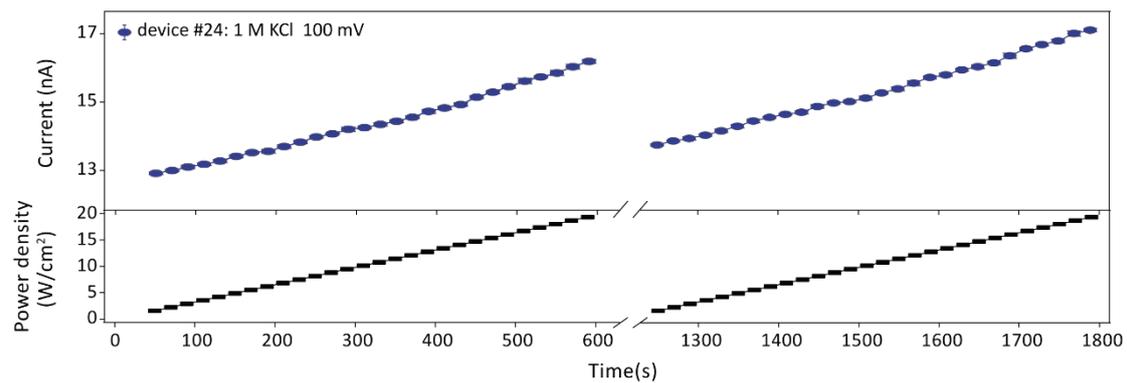

**Supplementary Figure 15. Ionic current optoelectronic modulation in a large MoS$_2$ pore.** Ionic current as a function of time in a 15 nm pore, recorded at fixed bias of 100 mV in 1 M KCl with two cycles, when light power density changes as a function of time (device #24).

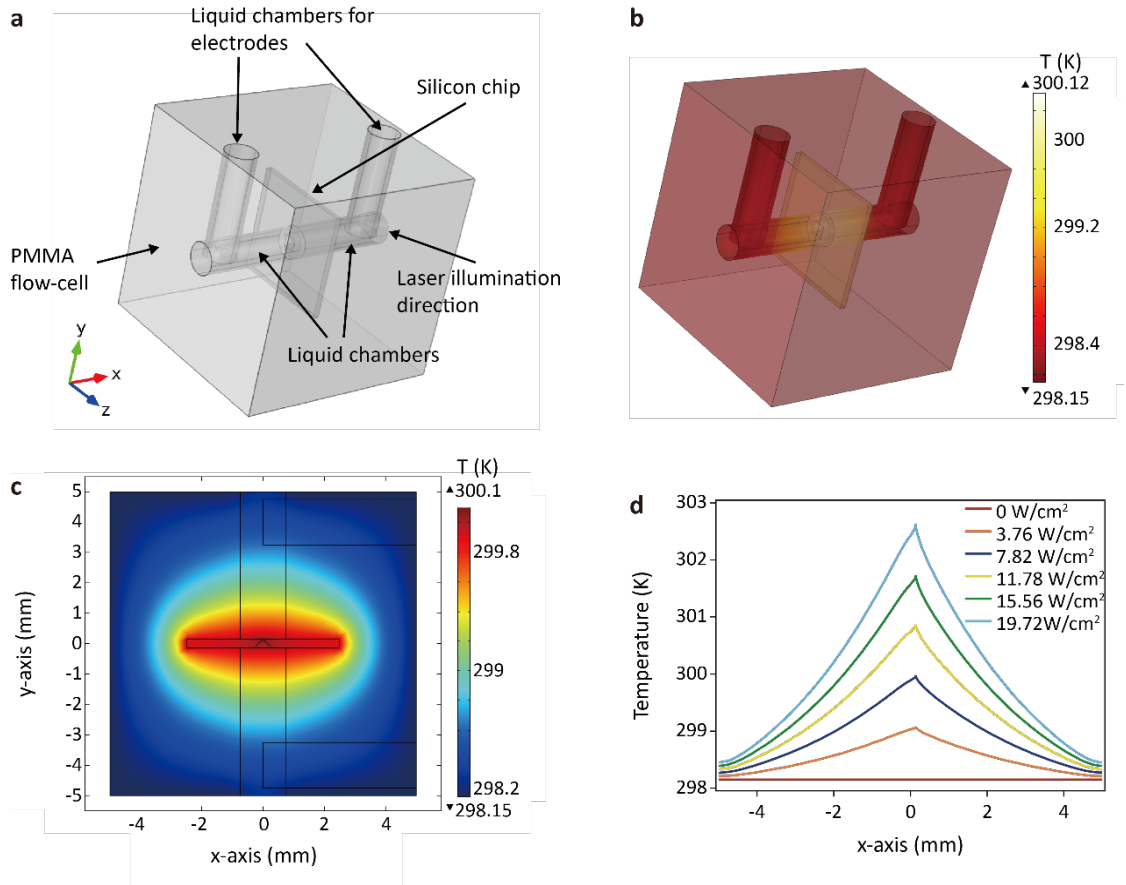

**Supplementary Figure 16. FEM model used to estimate the thermal effect of laser on the system. a.** Geometry of the used model. **b**. 3D surface heat map of the model. All surfaces of PMMA flow-cell were set to be 298.15 K as the boundary condition. Since this boundary condition is the most conservative estimate (the constant temperature surface may be closer to the silicon chip in fact), the increase for the practical system could be lower than the reported 4.5 K temperature changes. **c**. Heat map of the y-x axis of the system. **d**. The temperature profile along the x-axis at different laser power density.

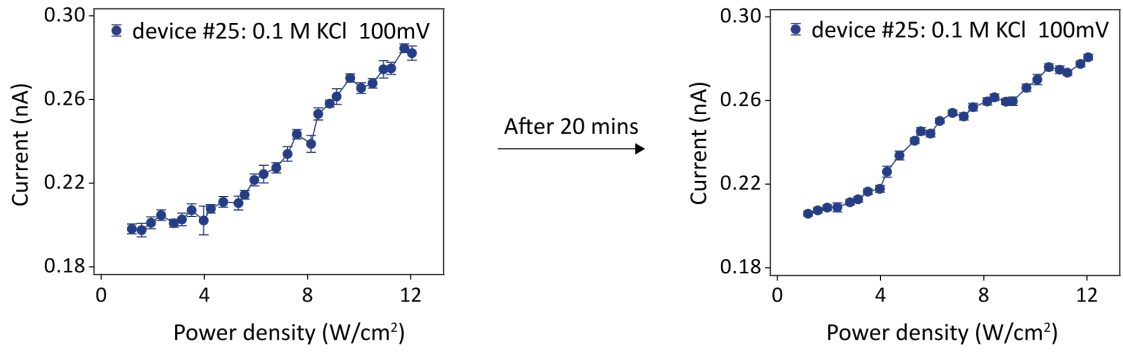

**Supplementary Figure 17. Ionic current modulation under 785 nm laser illumination.** Repeated linear response of current as a function of applied light power density with a 785 nm laser source, recorded at fixed bias of 100 mV in 0.1 M KCl (device #25: 0.7 nm).

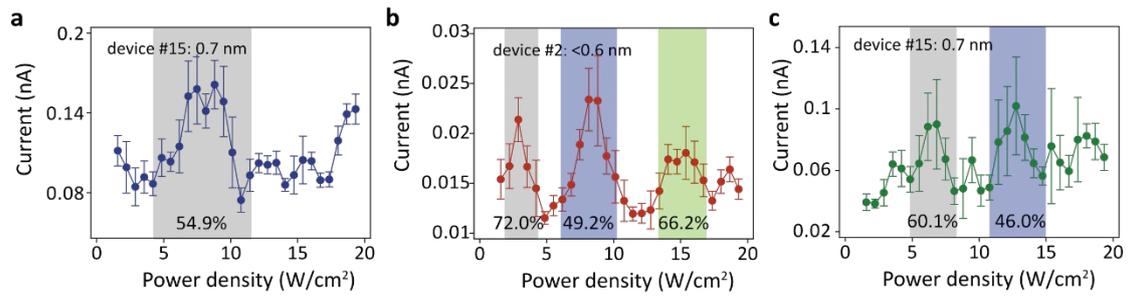

**Supplementary Figure 18. Ionic conduction oscillations in sub-nm MoS$_2$ pores. a-c.** Represented oscillatory results from 3 different sub-nm MoS$_2$ pore devices clearly show the patterns are distinguishable from the baseline noise, recorded from 3 different devices at a fixed bias of 400 mV in 0.1 M KCl, 500 mV in 0.05 M MgCl$_2$, 700 mV in 0.033 M AlCl$_3$ respectively.

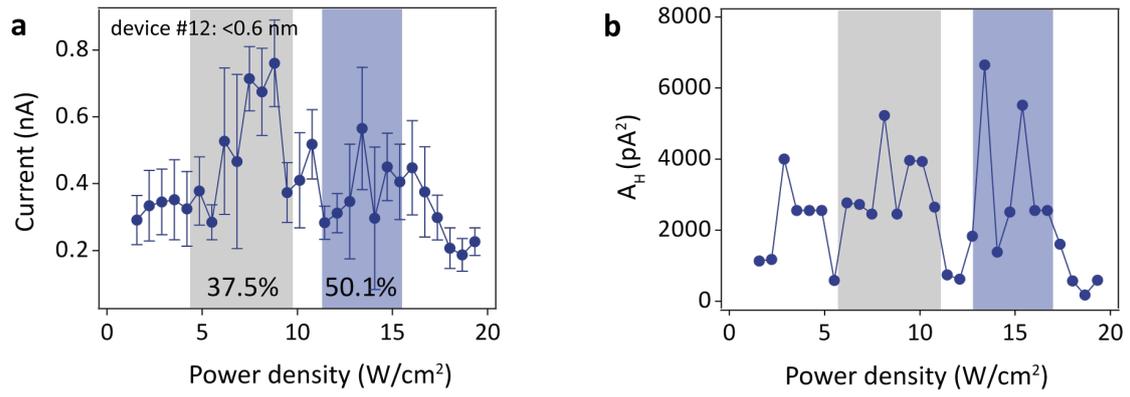

**Supplementary Figure 19. Noise spectral analysis of ionic conduction oscillations. a.** Ionic current oscillations recorded at a fixed bias of 600 mV in 0.1 M KCl (device #12: <0.6 nm). **b.** Evolution of the corresponding noise parameter $A_H$ with light power density modulation.

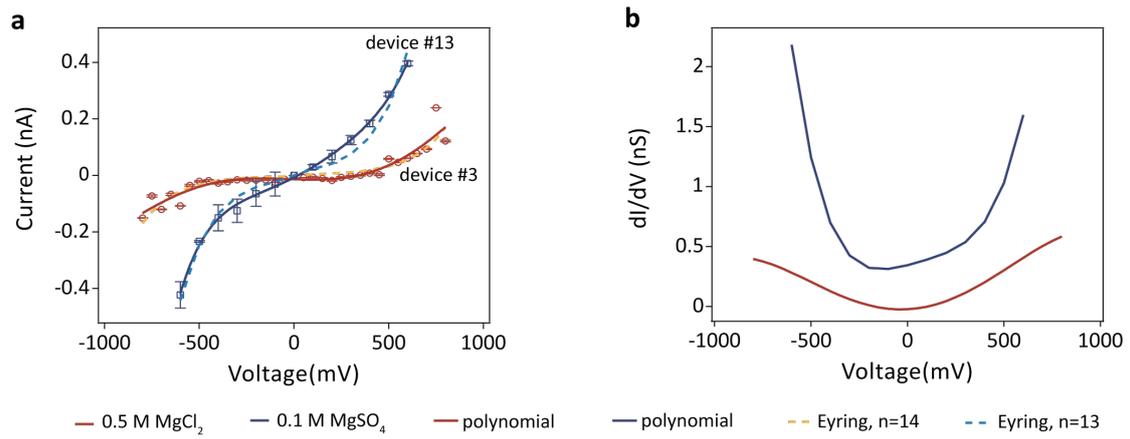

**Supplementary Figure 20.** *I-Vs* **and differential conductance for** $Mg^{2+}$. **a.** *I-V* characteristics of a sub-nm pore in 0.5 M $MgCl_2$ and 0.1 M $MgSO_4$ aqueous solutions at room temperature conditions. **b.** The corresponding differential conductance *dI/dV vs V* plot for the *I-V* data shown in **a**.

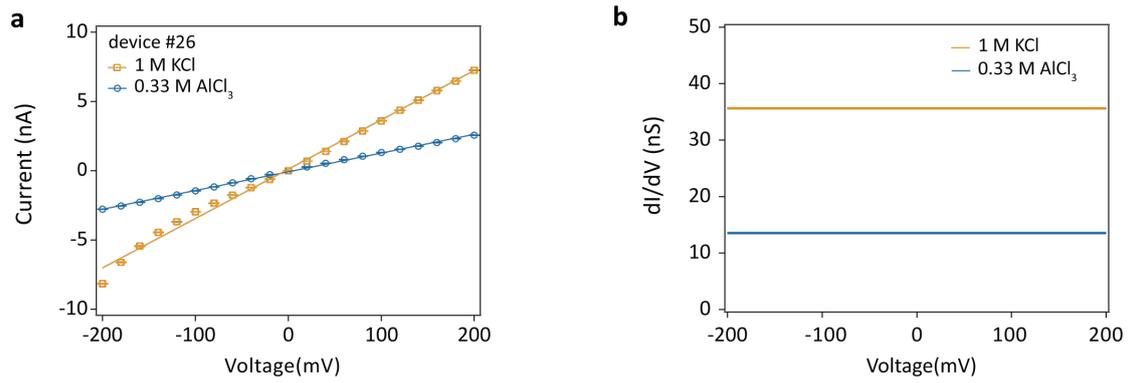

**Supplementary Figure 21.** *I-Vs* **and differential conductance comparison in 4 nm nanopore. a.** Linear *I-V* characteristics of a 4 nm $MoS_2$ pore in 1 M KCl and 0.33 M $AlCl_3$ aqueous solutions at room temperature conditions (device #26). **b.** The corresponding conductance *dI/dV vs V* plot for the *I-V* data shown in **a**.

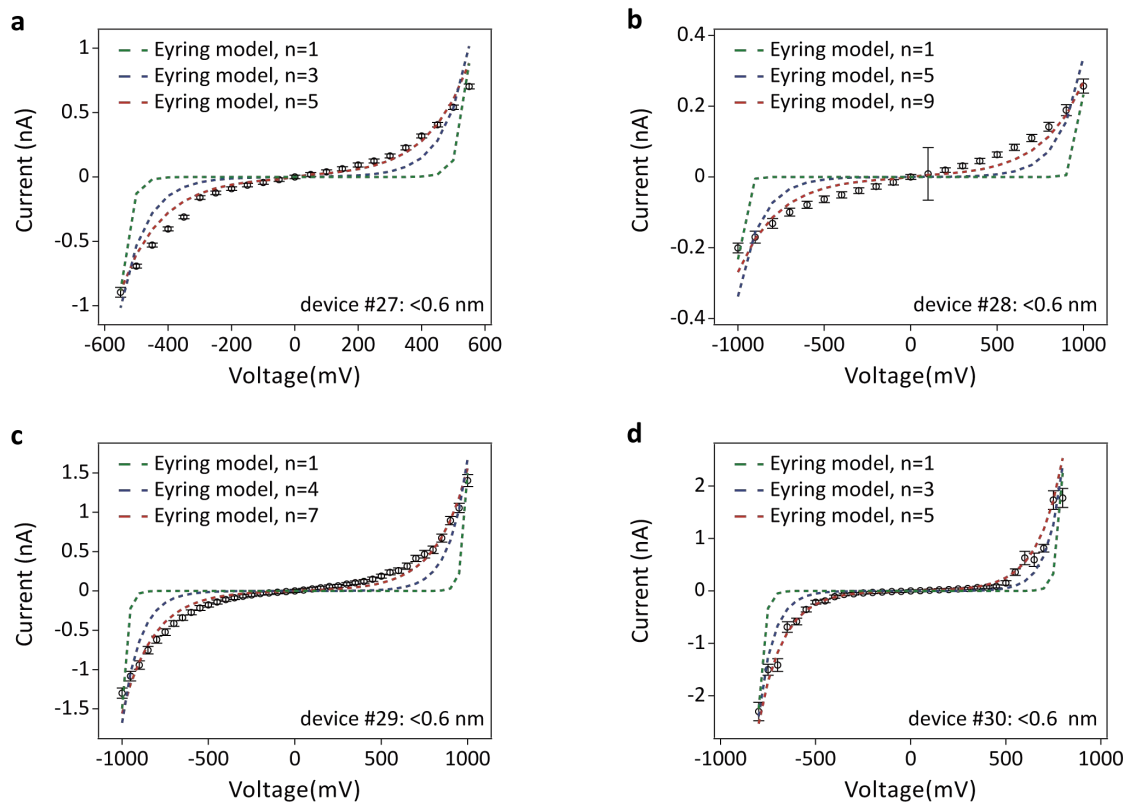

**Supplementary Figure 22. Eyring rate model applied to nonlinear *I-V*s. a-d.** Eyring model applied to examples of experimentally obtained *I-V*s from 4 different sub-nm $MoS_2$ pore devices in 1 M KCl. The fitting results suggest there are 5-7 barriers for ion conduction process in these systems.

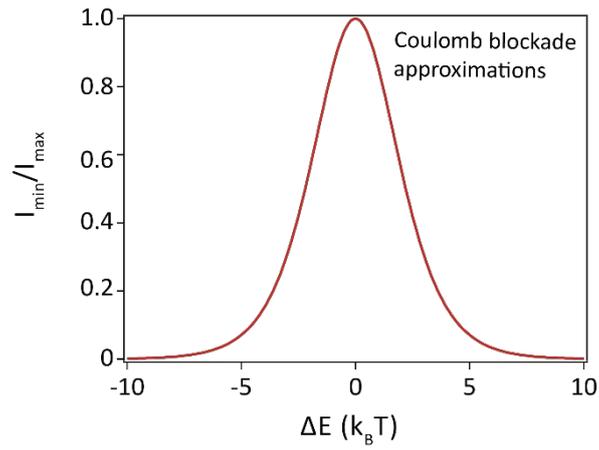

**Supplementary Figure 23. Theoretical analysis on the current oscillating ratio ($I_{min}/I_{max}$) vs. energy difference.** Data plot in figure is based on the Coulomb blockade oscillation approximation[34] at room temperature regime.

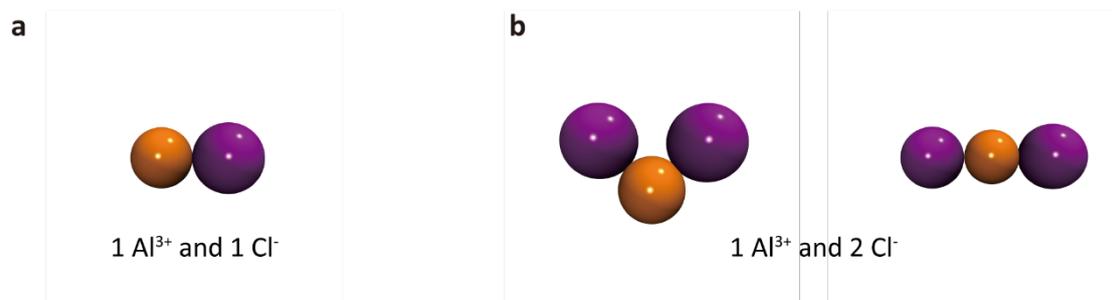

**Supplementary Figure 24. Ion pairs with aluminum ions and chloride ions observed in molecular dynamics simulations. a-b.** the ion pairs of aluminum ions and one chloride ion and two chloride ions. The simulation condition is 0.33 M $AlCl_3$ solution

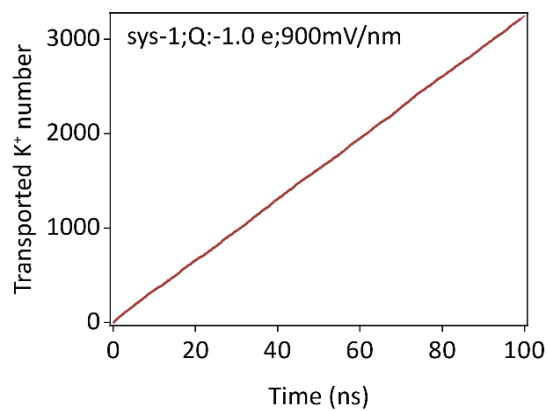

**Supplementary Figure 25. Extended simulation of a trajectory.** The trajectory of sys-1 with a bias voltage of 900 mV/nm is extended to 10 times length (up to 100 ns). The transported $K^+$ number reflects a good linear relation throughout the entire trajectory, which confirms 10 ns trajectory is enough and suitable to assess the transport characteristics of ions across nanopores.

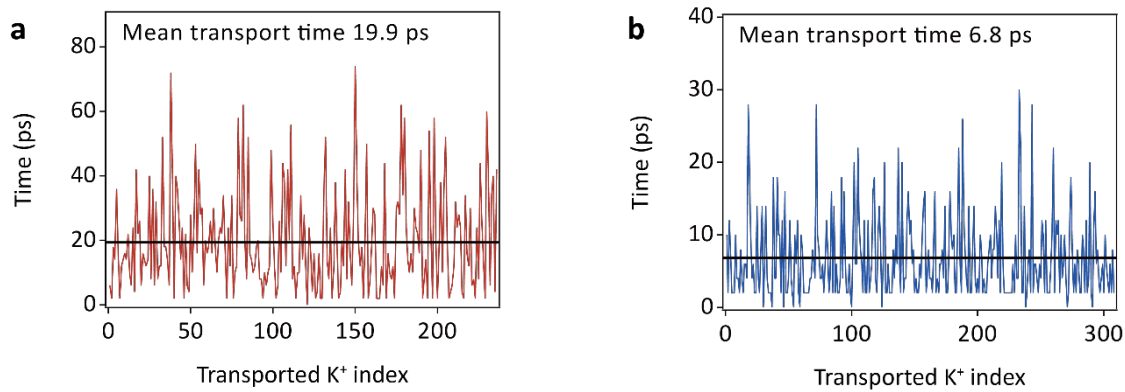

**Supplementary Figure 26. The comparison of transported times of K$^+$ ions.** The time is taken by each K$^+$ ion to pass through the MoS$_2$ nanopore in sys-1 under 900 mV/nm with nanopore active edge atoms charged by 0.0 e (**a**) and -0.5 e (**b**). The comparison confirms that K$^+$ is much more difficult to pass through the non-charged nanopore than through the charged nanopore, reflecting less transported number and longer passage time, which is well agreed with the PMF in main **Figure 4d**.

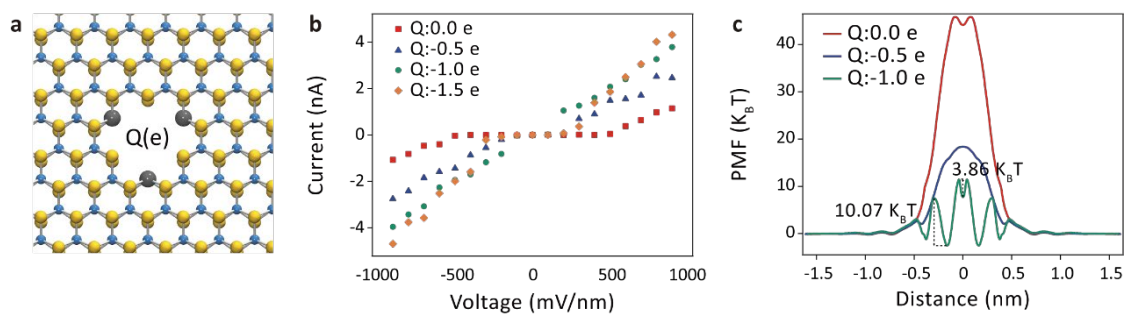

**Supplementary Figure 27. Molecular dynamics simulation of a hexagonal MoS$_2$ pore. a.** Snapshot of a hexagonal MoS$_2$ nanopore. **b.** Nonlinear *I-V* characteristics of MoS$_2$ nanopores by changing the molybdenum atoms' charges (0e, -0.5e, -1.0e and -1.5e) as denoted in **a**. **c.** Free energy profiles of a K$^+$ passing through sub-nm pores with different boundary charges. The charges *Q* (0 e, -0.5e and -1.0e) are separately endowed to the atoms in the pore boundary as denoted in **a**.

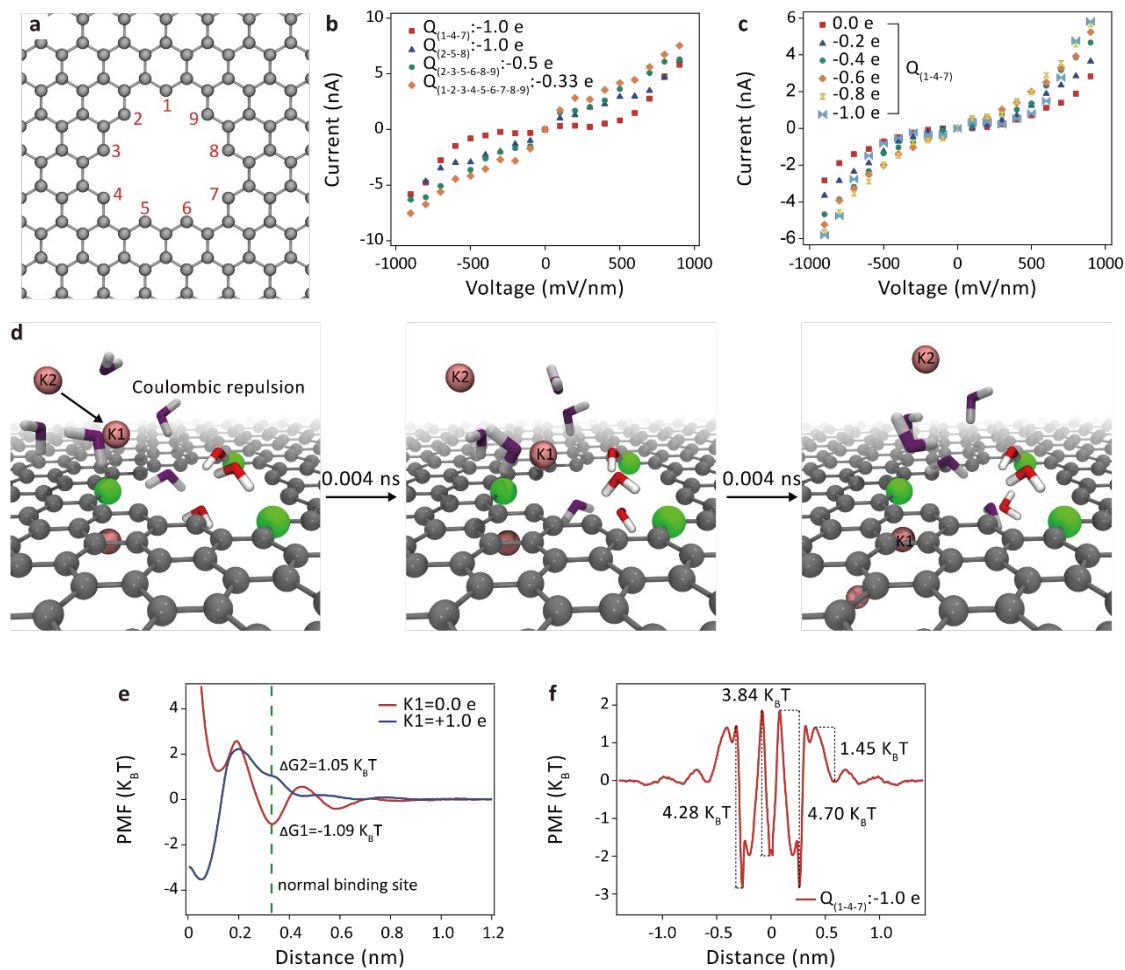

**Supplementary Figure 28. Molecular dynamics simulation of a graphene pore. a.** Snapshot of a sub-nm graphene nanopore. This nanopore has 9 carbon atoms near the pore, which are defined as 1 to 9. **b.** Nonlinear *I-V* characteristics of graphene pore with different pore charge. The charges are separately endowed to the atoms as marked. **c.** Nonlinear *I-V* characteristics of graphene nanopore with 1, 4 and 7 carbon atoms having different charges. **d.** Schematic of a potassium ion passing through $MoS_2$ nanopore caused by CB oscillation**.** The two key ions are shown with pink spheres, which are defined as K1 and K2. The 1, 4 and 7 carbon atoms are highlighted with green spheres. The water molecules near the pore are shown with red and white sticks while the water molecules near the K1 are shown with purple and white sticks. **e.** The PMFs of K1 (0e and +1.0e) trapped in the nanopore. Specifically, the PMFs evaluate the free energy alterations by fixing K1 ion while pulling K2 ion along the direction perpendicular to graphene, according to the positions of two ions in the first picture of **d**. The reaction coordinate (distance) indicates the distance between K1 and K2 ions along the direction perpendicular to graphene. **f.** Free energy profile of a $K^+$ passing through the graphene pore. The charge (-1.0e) is separately endowed to the 1, 4 and 7 carbon atoms in the pore boundary. And other atoms have zero charge. The reactive coordinate (distance) indicates the vertical interval of the $K^+$ and the nanopore center.

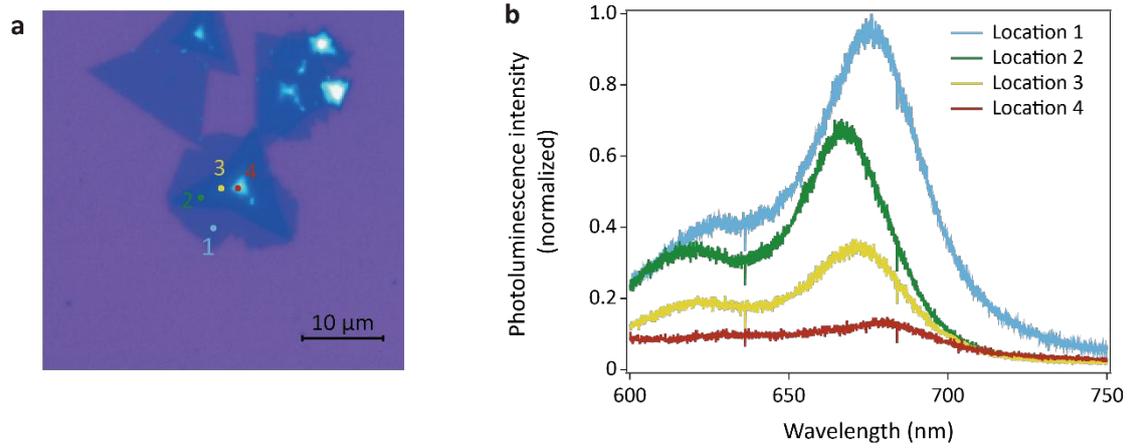

**Supplementary Figure 29. Photoluminescence measurements on MoS$_2$. a.** Optical image of CVD grown MoS$_2$ samples on Si/SiO$_2$ substrate. **b.** Photoluminescence results of the samples in **a**, marked with different MoS$_2$ thickness. For device fabrication, only high quality monolayer MoS$_2$ is used.

ix. **Supplementary Movies**

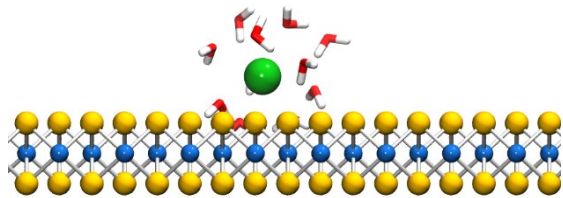

**Supplementary Movie 1.** A representative transport process of a single K$^+$ across the non-charged MoS$_2$ nanopore ($Q$: 0 e). The MoS$_2$ is displayed with orange spheres (S), light blue spheres (Mo) and white sticks (Mo-S bonds). The turquoise sphere indicates a K$^+$ ion. Water molecules near the K$^+$ ion (within 3.5 Å) at the initial frame of the video are shown with red and white sticks.

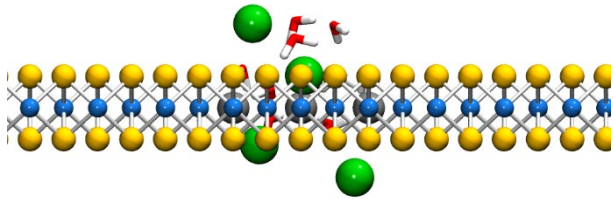

**Supplementary Movie 2.** A representative transport process of K$^+$ translocations through the charged MoS$_2$ nanopore ($Q$: -0.5e) with ion-ion interactions. This passage process is according to the main **Figure 4b**. The MoS$_2$ is displayed with orange spheres (S), light blue spheres (Mo) and white sticks (Mo-S bonds). The turquoise spheres indicate K$^+$ ions. Water molecules near the transported K$^+$ ion (within 3.5 Å) at the initial frame of the video are shown with red and white sticks.

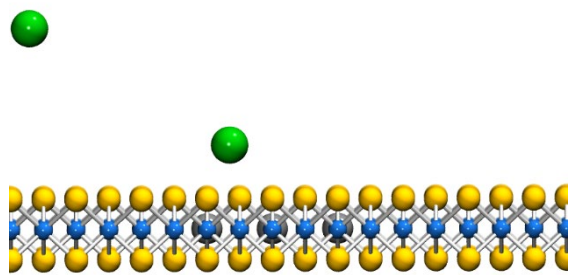

**Supplementary Movie 3.** A representative process of multi-ion (e.g., three $K^+$) in $MoS_2$ nanopore.

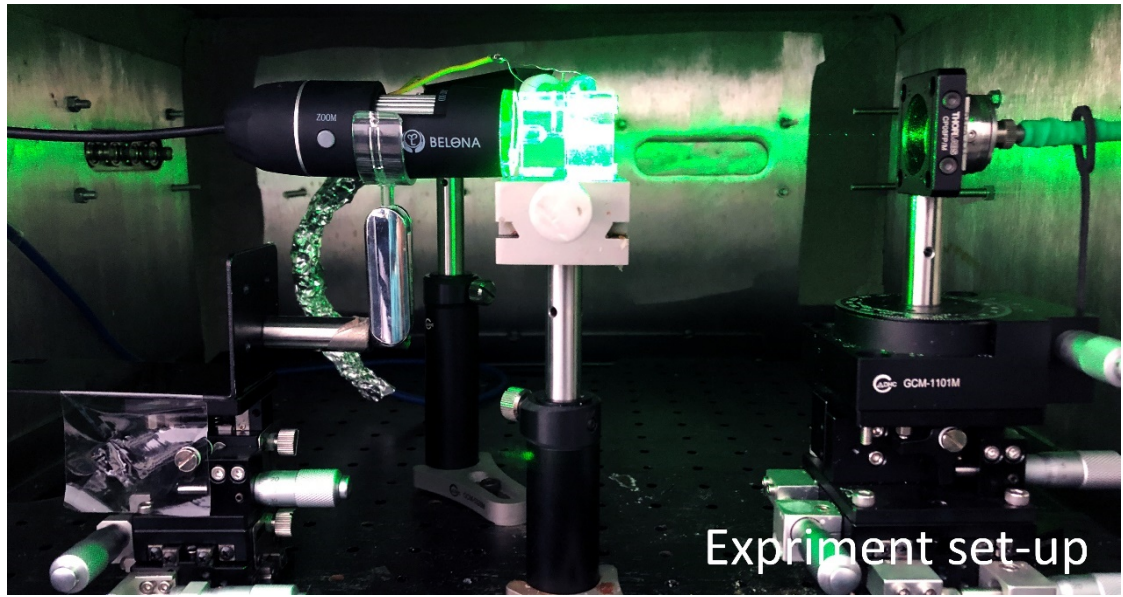

**Supplementary Movie 4.** The alignment of laser beam to the center of membrane. Details of beam properties are discussed in the methods and **Supplementary Figure 3**.